\documentclass[aps,pra,twocolumn,a4paper,floatfix,
superscriptaddress,preprintnumbers]{revtex4-1}
  \usepackage[T1]{fontenc}
  \usepackage{amsmath}
  \usepackage{amsbsy}
  \usepackage{amssymb}
\usepackage[utf8]{inputenc}
\usepackage[english]{babel}
\usepackage{ulem}

\usepackage{graphicx, color}
\newcommand{\D}{{\rm{d}}}
\newcommand{\I}{{\rm{I}}}
\newcommand{\Q}{{\rm{Q}}}

\begin{document}
\preprint{PHYSICAL REVIEW LETTERS {\bf 117}, 090501 (2016)}

\title{Atmospheric Quantum Channels with Weak and Strong Turbulence}

\author{D. Vasylyev}
\affiliation{Institut f\"ur Physik, Universit\"at Rostock,
Albert-Einstein-Stra\ss{}e 23, D-18059 Rostock, Germany}
\affiliation{Bogolyubov Institute for Theoretical Physics, NAS of Ukraine, Vulytsya
Metrologichna 14-b, 03680 Kiev, Ukraine}

\author{A. A. Semenov}

\affiliation{Institut f\"ur Physik, Universit\"at Rostock,
Albert-Einstein-Stra\ss{}e 23, D-18059 Rostock, Germany}
\affiliation{Institute of Physics, NAS of Ukraine, Prospect Nauky 46, 03028
Kiev, Ukraine}

\author{W. Vogel}
\affiliation{Institut f\"ur Physik, Universit\"at Rostock,
Albert-Einstein-Stra\ss{}e 23, D-18059 Rostock, Germany}

\begin{abstract}
The free-space transfer of high-fidelity optical signals  between remote locations has many applications, including both
classical and quantum communication, precision navigation, clock synchronization, etc.  The physical processes
that contribute to  signal fading and loss  need to be carefully analyzed in the theory of light propagation through the atmospheric
turbulence. Here we derive the probability distribution for the atmospheric
transmittance including beam wandering, beam shape deformation, and
beam-broadening effects. Our model, referred to as the elliptic-beam approximation,
applies to weak, weak-to-moderate, and strong turbulence and hence to
the most important regimes in atmospheric communication scenarios.
\end{abstract}

\maketitle

\paragraph*{Introduction.--} The transmission of quantum light to remote receivers 
recently attracted great interest in connection with the implementation of quantum 
communication protocols over large distances.
Experimental advances in this field allowed one to demonstrate the successful quantum-light transmission  over horizontal communication links ~\cite{Ursin, Scheidl, Fedrizzi2009, Capraro, Yin, Ma, Peuntinger}
and paved the way for the  realization of ground-to-satellite quantum
communication~\cite{Nauerth, Wang, Bourgoin}.
The main obstacle for the transmission of quantum light in free space is the
atmospheric turbulence, which  leads to  spatial and temporal  variations of the
refractive index of the channel. The transmitted signal
is usually measured by detectors with a finite aperture. Typically, the recorded
data are contaminated by fluctuating losses due to beam wandering, beam
broadening, scintillation, and degradation of coherence.

The theory of classical light propagation through the atmosphere is well
developed~\cite{Tatarskii, Ishimaru, Andrews, Andrews2, Fante1, Fante2}. Some progress
was also achieved in
the
theory of  free-space propagation of quantum light \cite{Diament,Perina,
Perina1973, Milonni, Paterson, Semenov, Semenov10, Vasylyev2012}.
The atmosphere is
considered as a quantum channel characterized by  fluctuating transmission properties.
In terms of the Glauber-Sudarshan
$P$ function~\cite{Glauber, GlauberPRA, Sudarshan}, which is a quasiprobability as
it may
attain negativities, the relation between input
$P_\mathrm{in}\!\left(\alpha\right)$ and
output $P_\mathrm{out}\!\left(\alpha\right)$ states can be written
as~\cite{Semenov,Vasylyev2012}
\begin{align}
P_{\rm out}(\alpha)=\int\limits_0^1 \D \eta\,\mathcal{P}(\eta)\frac{1}{\eta}P_{\rm
in}\Bigl(\frac{\alpha}{\sqrt{\eta}}\Bigr).\label{inout}
\end{align}
Here $\mathcal{P}(\eta)$ is the probability distribution of the transmittance (PDT), $\eta$
being the intensity transmittance.
Hence, the description of quantum-light propagation through the turbulent atmosphere
merely reduces to identifying this probability distribution. In Ref.~\cite{Vasylyev2012} we have
derived the PDT for the case when the leading effect of fluctuating losses in the
atmosphere is beam wandering, as it is the case for weak turbulence.

In this Letter, we present a substantially extended model of PDT, based on the
{\it elliptic-beam approximation} that incorporates effects of beam wandering,
broadening, and deformation.  Our theory properly
describes atmospheric quantum channels in the limits of relatively weak
turbulence, as in experiments
in Erlangen with an atmospheric link of 1.6 km
length~\cite{Usenko,Peuntinger}. For the case of strong turbulence, our theory also 
yields a reasonable agreement with the log-normal
model~\cite{Andrews,Andrews2, Fante1, Fante2, Diament, Perina,Perina1973,
Milonni}, which has been verified in experiments on the Canary
Islands~\cite{Capraro}. Most importantly, our elliptic-beam model overcomes the
deficiency of physical inconsistencies inherent in the log-normal
distribution.

\paragraph*{The aperture transmittance.--}
Temporal and spatial variations of  temperature and pressure in atmospheric
turbulent flows cause random fluctuations of the refraction index of the air.
Consequently, the atmosphere acts as a source of losses for transmitted photons which
are measured  at the receiver by a detection module with a finite aperture. The transmitted signal is
degraded by effects like beam wandering, broadening, deformation, and others.
Let us consider a Gaussian beam that propagates along
the $z$~axis onto the aperture plane at distance $z=L$. In general, the
fluctuating intensity transmittance  of such a signal  is given by
\cite{Vasylyev2012}
\begin{align}
 \eta&{=}\int_{\mathcal A}\D^2 \mathbf{r}\, I(\mathbf{r};L),\label{Transm}
\end{align}
where $\mathcal{A}$ is the aperture area and $I(\mathbf{r};L)$ is the
normalized intensity with respect to the full $\textbf{r}{=}\{x,y\}$ plane.

The Gaussian beam underlies turbulent disturbances along the
propagation path. Within our model we assume that
these disturbances lead to beam wandering and deformation of the beam profile
into an elliptical form. This is justified for weak turbulence, when speckles
play no essential role. For strong turbulence the beam shape is the result of
many small spatially averaged distortions.
The intensity of the elliptic beam at the aperture plane
is given by
\begin{align}
 I(\mathbf{r};L)=\frac{2}{\pi\sqrt{\det\mathbf{S}}}\exp\Bigl[-2({\mathbf{r}}{-}{
\mathbf{r}}_0)^{\rm
T}{\mathbf{S}}^{-1}({\bf r}{-}{\bf r}_0)\Bigr],\label{Intens}
\end{align}
with $\mathbf{r}{=}(x\,\,y)^{\rm T}$.
It is characterized by  the beam-centroid position $\mathbf{r}_0=(x_0\,\,y_0)^{\rm T}$ and
the real, symmetric, positive-definite spot-shape matrix $\mathbf{S}$. The
eigenvalues of this matrix,
$W_i^2$,  $i{=}1,2$, are squared semiaxes of the elliptic spot.
The semiaxis $W_1$ has an angle $\phi{\in}\left[0,\pi/2\right)$ relative to the $x$ axis, and
the set $\left\{W_1^2,W_2^2,\phi\right\}$ uniquely describes the
orientation and the size of the ellipse.

For an elliptic-beam profile, the transmittance $\eta$ is obtained by substituting Eq.~(\ref{Intens})
into Eq.~(\ref{Transm}). The resulting integral cannot be
evaluated analytically. Here we adapt the technique proposed in
Ref.~\cite{Vasylyev2012} to derive an analytical
approximation. For this purpose we consider the displacement of the beam centroid to
the point
$\mathbf{r}_0{=}\left(r_0\cos\varphi_0\,\,\,r_0\sin\varphi_0\right)^\textrm{T}$.
Regarding the transmittance $\eta$ as a function of $r_0$, for given
$\chi{=}\phi{-}\varphi_0$, we observe that it behaves similar to the
transmittance of the circular Gaussian beam with the effective squared spot radius
\begin{align}
 W_\textrm{eff}^2\left(\chi\right)&{=}4a^2\Bigl[\mathcal{W}\Bigl(\frac{4a^2}{
W_1W_2} e^
{\frac{a^2}{W_1^2}\bigl\{1+2\cos^2\!\chi\bigr\}}\nonumber\\
 &\qquad\times
e^
{\frac{a^2}{W_2^2}\bigl\{1+2\sin^2\!\chi\bigr\}}\Bigl)\Bigr]^{-1},\label{Weff}
\end{align}
where $\mathcal{W}(\xi)$ is the Lambert $W$ function \cite{Corless} and $a$ is the
aperture radius.
In this case the transmittance is approximated by
\begin{align}
 \eta=\eta_{0} \exp\left\{-\left[\frac{r_0/a}
 {R\left(\frac{2}{W_{\rm
eff}\left(\phi{-}\varphi_0\right)}\right)}\right]^{\lambda\bigl(\frac{2}{W_{\rm
eff}\left(\phi{-}\varphi_0\right)}\bigr)}\right\}.\label{Tapprox}
\end{align}
Here $\eta_0$ is the transmittance for the centered beam, i.e. for
$r_0{=}0$,
\begin{align}
 &\eta_{0}{=}1{-}\I_0\Bigl(a^2\Bigl[\frac{1}{W_1^2}{-}\frac{1}{W_2^2}\Bigr]\Bigr)e^{-a^2\bigl[\frac{1}{W_1^2}{+}\frac{1}{W_2^2}\bigr]}\nonumber\\
 &\qquad{-}2\left[1{-}e^{-\frac{a^2}{2}\!\bigl(\frac{1}{W_1}{-}\frac{1}{W_2}\bigr)^{2}}\!\right]\nonumber\\
 &\qquad\times\exp\!\left\{\!{-}\Biggl[\!\frac{\frac{(W_1+W_2)^2}{|W_1^2-W_2^2|}}{R\left(\frac{1}{W_1}{-}\frac{1}{W_2}\right)}\!\Biggr]
 ^{\!\lambda\left(\!\frac{1}{W_1}{-}\frac{1}{W_2}\right)}\right\},
\end{align}
$R(\xi)$ and $\lambda(\xi)$ are  scale and shape functions, respectively,
\begin{align}
 R\left(\xi\right)=\Bigl[\ln\Bigl(2\frac{1-\exp[-\frac{1}{2} a^2
\xi^2]}{1-\exp[-a^2\xi^2]\I_0\bigl(a^2\xi^2\bigr)}\Bigr)\Bigr]^{-\frac{1}{
\lambda(\xi)}},
\end{align}
\begin{align}
 \lambda\left(\xi\right)&=2a^2\xi^2\frac{e^{-a^2\xi^2}\I_1(a^2\xi^2)}{1-\exp[
-a^2\xi^2 ] \I_0\bigl(a^2\xi^2\bigr)}\nonumber\\
 &{\times}\Bigl[\ln\Bigl(2\frac{1-\exp[-\frac{1}{2} a^2 \xi^2]}{1-\exp[-a^2\xi^2]\I_0\bigl(a^2\xi^2\bigr)}\Bigr)\Bigr]^{-1},
\end{align}
and $\I_{i}(\xi)$ is the modified Bessel function of $i$th order. Since $\phi$ is defined by
modulo $\pi/2$, the transmittance $\eta$ is a $\pi/2$-periodical function of $\phi$.
For the
limit $W_1^2{=}W_2^2$, Eq.~(\ref{Tapprox}) reduces to the transmission
coefficient
of a Gaussian beam with a circular profile~\cite{Vasylyev2012}. For details of
the approximation see Supplemental Material~\cite{suppl} and Ref.~\cite{Agrest}.

\paragraph*{The probability distribution of the transmittance.--} The
aperture transmittance $\eta$, cf.~Eq.~(\ref{Tapprox}),
is a
function of five real parameters, $\left\{x_0,y_0,\Theta_1,\Theta_2,\phi\right\}$,
randomly changed by the atmosphere, where $W_{i}^2{=}W_0^2\exp\Theta_{i}$, and
$W_0$ is the initial beam-spot radius. For these parameters we assume a Gaussian
approximation,
with $\phi$ being a
$\pi/2$-periodical wrapped Gaussian variable~\cite{Mardia}. We restrict our attention to
isotropic turbulence. In this case the wrapped Gaussian distribution for $\phi$ reduces to
a uniform one and its correlations with other parameters vanishes. In the reference frame
with $\big\langle\textbf{r}_0\big\rangle{=}0$, there are also no correlations of $x_0$,
$y_0$, and $\Theta_i$.

The variances $\langle\Delta x_0^2\rangle=\langle\Delta
y_0^2\rangle=\langle x_0^2\rangle$, which describe beam wandering, are expressed
in terms of the classical field correlation function of the fourth order,
$\Gamma_{4}(\mathbf{r}_1,\mathbf{r}_2)=\langle I(\mathbf{r}_1;L) I
(\mathbf{r}_2;L)\rangle$, in the aperture plane (see, e.g.,~\cite{Andrews, Andrews2, Fante1,
	Fante2, Mironov, Banakh, Kon, Chumak, Mironov2}):
\begin{align}
\langle x_0^2\rangle=\int_{\mathbb{R}^4} \D^4 \mathbf{r}\,
x_1 x_2
\,
\Gamma_4(\mathbf{r}_1,\mathbf{r}_2),\label{bwvariance}
\end{align}
where $\D^4\mathbf{r}=\D^2\mathbf{r}_1\D^2\mathbf{r}_2$. The means and the
(co)variances of $\Theta_i$ are functions of the means and the (co)variances (first and
second moments) of $W_i^2$:
\begin{align}
\langle
\Theta_{i}\rangle=\ln\left[\frac{\langle
	W_{i}^2\rangle}{W_0^2}\left(1+
\frac{\langle (\Delta	W_{i}^2)^2\rangle}{\langle	
	W_{i}^2\rangle^2}\right)^{-1/2}\right],\label{Eq:ThetaMean}
\end{align}
\begin{align}
\langle \Delta\Theta_i\Delta\Theta_j\rangle=
\ln\left[1+
\frac{\langle \Delta W_i^2 \Delta W_j^2\rangle}{\langle	
	W_i^2\rangle\langle	
	W_j^2\rangle}\right].\label{Eq:ThetaCovariances}
\end{align}
In general,  the evaluation of $\langle W_{i}^2\rangle$ and $\langle\Delta W_{i}^2\Delta W_{j}^2\rangle$ 
in Eqs.~(\ref{Eq:ThetaMean}) and (\ref{Eq:ThetaCovariances}) is almost intractable.
However the
assumptions of Gaussianity and isotropy enable to express these quantities
in a  tractable form as (for details cf.~the Supplemental Material~\cite{suppl})
\begin{align}
	\langle&
	W_{i}^2\rangle{=}4\left[\int_{\mathbb{R}^2}\D^2\mathbf{r}\, x^2 \Gamma_2\!
	\left(\mathbf{r}\right){-}
	\langle x_0^2\rangle\right],\label{Eq:W12SqMean}
\end{align}
\begin{align}
&\langle
W_{i}^2W_{j}^2\rangle=8\Big[{-}8\,\delta_{ij}\langle
x_0^2\rangle^2
{-}\langle x_0^2\rangle \langle
W_{i}^2\rangle
\label{Eq:W12ViaGamma}\\
&+\int_{\mathbb{R}^4}\D^4\mathbf{r}
\,
\left[x_1^2x_2^2\left(4\delta_{ij}{-}1\right)-x_1^2y_2^2\left(4\delta_{ij}{-}3\right)\right]
\,\Gamma_4\!\left(\mathbf{r}_1,\mathbf{r}_2\right)
\Big],\nonumber
\end{align}
where $\Gamma_{2}(\mathbf{r}){=}\langle I(\mathbf{r};L)\rangle$ is the classical field
correlation function of the second order. 

Therefore, the means and the covariance matrix of the random vector
$\textbf{v}{=}\big(x_0\,y_0\,\Theta_{1}\,\Theta_{2}\big)^\mathrm{T}$,
i.e. $\mu_i{=}\left\langle v_i\right\rangle$ and
$\Sigma_{ij}{=}\left\langle\Delta v_i\Delta v_j\right\rangle$, respectively, are expressed in
terms of classical field correlation functions $\Gamma_2$ and $\Gamma_4$. These
functions are important characteristics of atmospheric channels, which are widely
discussed in the literature; see, e.g., \cite{Andrews, Andrews2, Fante1, Fante2,Mironov,
Banakh, Mironov2}. In the Supplemental Material~\cite{suppl}, we derive $\mu_i$ and $\Sigma_{ij}$
for horizontal links by using the phase approximation of the Huygens-Kirchhoff method
and the Kolmogorov turbulence spectrum~\cite{Mironov, Banakh,Mironov2}.

With the given assumptions, the PDT in Eq.~(\ref{inout}) reads as
\begin{align}
 \mathcal{P}\left(\eta\right){=}\frac{2}{\pi}\int_{\mathbb{R}^4}\D^4\mathbf{v}
\int\limits_{0}^{ \pi /2}\D\phi \,
\rho_G(\mathbf{v};\boldsymbol{\mu},\Sigma)\delta\left[\eta{-}\eta\left(\mathbf{v},\phi\right)\right],\label{PDTC}
\end{align}
where $\eta\left(\mathbf{v},\phi\right)$ is the transmittance defined
by Eq.~(\ref{Tapprox}) as a function of random parameters and
$\rho_G(\mathbf{v};\boldsymbol{\mu},\Sigma)$ is the Gaussian probability density of the vector $\mathbf{v}$ with the mean  $\boldsymbol{\mu}$ and the covariance matrix $\Sigma$.
In general, the PDT can be evaluated with the Monte Carlo method.
For this purpose, one has to simulate the Gaussian random vector $\mathbf{v}$ and
the uniformly distributed angle $\phi$.
For practical purposes, we apply the Rayleigh distribution for $r_0$, a uniform distribution for $\chi$, \
and a Gaussian one for $\Theta_i$. The obtained values should be
substituted in the transmittance; cf.~Eq.~(\ref{Tapprox}). Within
the standard procedure of estimation, one obtains the mean value of any
function of the transmittance, $\langle f(\eta)\rangle$. The PDT can be obtained
within the smooth-kernel method~\cite{Wand}.  The cumulative probability
distribution, $\mathcal{F}(\eta){=}\int_0^{\eta}\D
\eta^\prime\mathcal{P}(\eta^\prime)$, and the exceedance
$\overline{\mathcal{F}}(\eta)=1-\mathcal{F}(\eta)$ are estimated by the technique of
empirical distribution functions~\cite{Vaart}.

\paragraph*{From weak to strong turbulence.--} Let us distinguish the regimes of
weak, moderate, and strong turbulence, through the values of   Rytov
parameter  $\sigma_{\rm R}^2<1$, $\sigma_{\rm R}^2\approx 1 \dots 10$, and
$\sigma_{\rm R}^2\gg 1$, respectively. The Rytov parameter is defined as
$\sigma_R^2{=}1.23 C_n^2k^{\frac{7}{6}}L^{\frac{11}{6}}$, where $C_n^2$ is the
atmospheric index-of-refraction structure constant and $k$ is the optical wave number; 
for more details and the corresponding motivation, see Ref.~\cite{Andrews}. For weak  turbulence  
the atmosphere mainly causes beam wandering. In this case Eq.~(\ref{PDTC}) 
reduces to the log-negative Weibull distribution~\cite{Vasylyev2012}.
For the weak-to-moderate transition and for strong turbulence, broadening and
deformation of the beam occur, resulting in a smooth PDT. More problematic
is the evaluation of $\mathbf{v}$ and $\Sigma$ for the moderate-to-strong 
turbulence transition. Hence we will restrict our considerations to the ranges of 
weak-to-moderate and strong turbulence.

\begin{figure}[h]
 \includegraphics{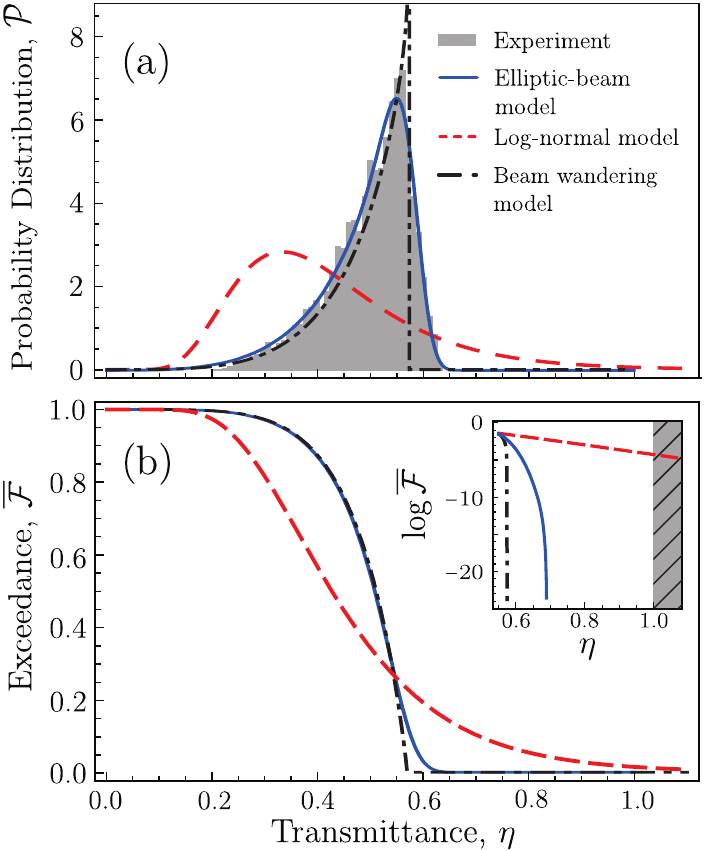}
 \caption{\label{fig:PDTC} The PDTs
 (a)
 and the corresponding exceedances  (b): elliptic-beam approximation,  log-normal,
 and beam wandering~\cite{Vasylyev2012}. The shaded
 area in (a) shows the experimental PDT from Ref.~\cite{Usenko}.
 The inset in (b) shows the
tail of the exceedance. For the
log-normal exceedance, it extends to the unphysical
 region, $\eta>1$ (shaded gray). Further parameters: wavelength $809\,\textrm{nm}$,  initial
spot radius $W_0{=}20\,\textrm{mm}$, propagation distance $1.6\,\textrm{km}$, Rytov parameter $\sigma_R^2{=}1.5$, aperture radius $a=40\,\textrm{mm}$, deterministic
 attenuation of $1.25\,\textrm{dB}$.}
\end{figure}

Figure~\ref{fig:PDTC} shows the probability
$\mathcal{P}(\eta)$ derived
by the elliptic-beam approximation for the conditions of weak-to-moderate
turbulence.
This distribution is compared with the
corresponding ones obtained  from the beam-wandering model~\cite{Vasylyev2012} and
from the log-normal model, see Supplemental Material~\cite{suppl}. The inset shows the  experimental data given in
Ref.~\cite{Usenko}. It is obvious that the elliptic model yields the best agreement with the measured data.

The log-normal distribution is quite popular for modeling atmospheric turbulence
effects~\cite{Andrews,Andrews2, Fante1, Fante2, Diament, Perina,Perina1973, Milonni}.
Usually this model
is applied for the description of intensity fluctuations in one spatial point. In Fig.~\ref{fig:PDTC},  the  log-normal model is applied to the signal detection with a finite aperture; see Supplemental Material~\cite{suppl}.
The dashed line in Fig.~\ref{fig:PDTC} (a) shows that
the log-normal distribution differs significantly from the measured PDT.
Moreover, the  log-normal PDT  is not  limited to the physically allowed interval $\eta\in[0,1]$.
This feature is clearly seen in
Fig.~\ref{fig:PDTC} (b) where the exceedance  functions
$\overline{\mathcal{F}}(\eta)$, i.e., the
probability that the transmittance  exceeds
the value of $\eta$, are shown for the elliptic model, the beam-wandering model,  and the
log-normal model. As was shown in~Ref.\cite{Semenov10,Vasylyev2012},
the tails of $\overline{\mathcal F}$
with large values of $\eta$ are important for preserving nonclassical properties
of transmitted light, which are overestimated by the log-normal model.

\begin{figure}[h]
 \includegraphics{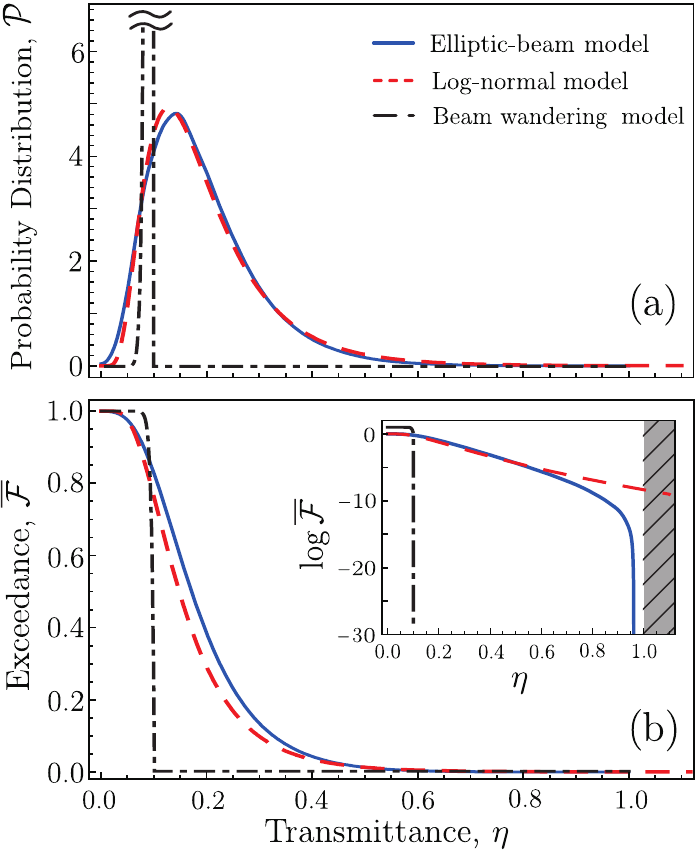}
 \caption{\label{fig:PDTC_StrongTurb} The PDTs and the
corresponding exceedances, similar to those shown in Fig.~\ref{fig:PDTC}, but for
the case of strong turbulence. Further parameters:
wavelength $780\,\textrm{nm}$,  initial
spot radius $W_0{=}50\textrm{mm}$, propagation distance
$2\,\textrm{km}$, Rytov parameter $\sigma_R^2{=}31.5$, aperture radius
$a=150\,\textrm{mm}$ and no deterministic attenuation.}
\end{figure}

It has been shown in experiments with
coherent light propagating through a $144\,\textrm{km}$ atmospheric channel on
the Canary Islands~\cite{Capraro} that the log-normal distribution in its physical
domain demonstrates a good agreement with the experimental data under the
conditions of strong turbulence. In Fig.~\ref{fig:PDTC_StrongTurb}, we compare
the PDTs derived from the elliptic-beam approximation with the ones obtained
in the beam-wandering and the log-normal models. Although in this case we consider
a short propagation distance, the turbulence is  quite strong. Similar
conditions may occur, e.g., for the case of near-to-ground  propagation on a hot summer day.

From Fig.~\ref{fig:PDTC_StrongTurb} one can clearly conclude that the beam-wandering
model strongly differs from the log-normal distribution and,
consequently, it cannot well describe the strong turbulence scenario. 
However, the elliptic-beam model gives a reasonable agreement with the 
log-normal distribution in the physical domain of the latter. 
From this fact, one may conclude that our model
consistently describes also the case of strong turbulence. A clear advantage
of the elliptic-beam model is that the corresponding PDT does not
attain nonzero values in the
unphysical domain, $\eta{>}1$, which is the case for the log-normal distribution.  
Hence, the usage of the elliptic-beam model gives physically consistent
results, whereas the log-normal distribution may yield unphysical artifacts, e.g., the
creation of photons by the atmosphere~\cite{Perina1973}. Such artifacts
may cause an overestimation of the security of quantum communication protocols. 
Finally, we note that in some cases beam wandering is suppressed by tracking procedures~\cite{Ursin, Nauerth}.
Under such conditions, the beam-wandering model is no longer useful but the elliptic-beam 
model does apply.

\paragraph*{Application: quadrature squeezing.--} The PDT (\ref{PDTC}) in
the
elliptic-beam approximation allows one to analyze the quantum properties of light
transmitted through the turbulent atmosphere by means of the input-output
relation~(\ref{inout}). As an example, we analyze the squeezing
properties after a weak-to-moderate turbulent atmospheric channel. We
consider the $1.6\,\textrm{km}$ link  in the city of Erlangen
\cite{Peuntinger}.
The transmitter generates  squeezed light ($-2.4\,\textrm{dB}$) at
$\lambda=780\,\textrm{nm}$ and sends it through the link with the Rytov parameter
$\sigma_R^2=2.6$.
The receiver  detects  $-0.95\,\textrm{dB}$ of squeezing.

\begin{figure}[ht!]
 \includegraphics{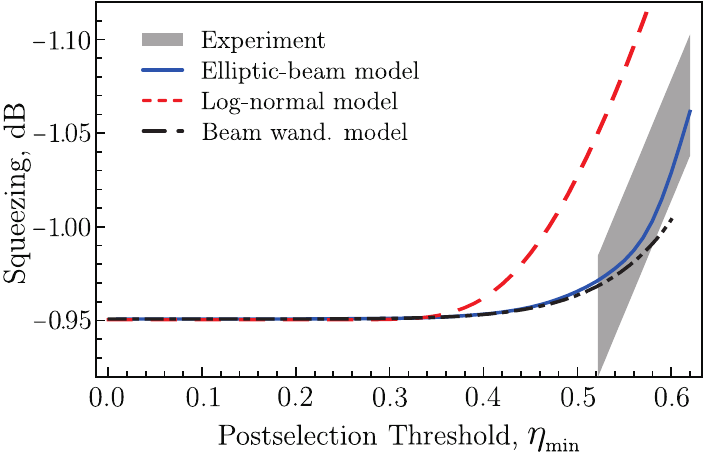}
 \caption{\label{fig:sq} Transmitted value of
 squeezing as a function of the postselection threshold $\eta_\textrm{min}$.
Initially squeezed light
(to $-2.4\,\textrm{dB}$, $\lambda=780\,\textrm{nm}$, spot radius
$W_0=25\,\textrm{mm}$)
 is sent through a $1.6\,\textrm{km}$
 atmospheric link ($\sigma_R^2=2.6$) and detected with an aperture radius of
 $a=75\,\textrm{mm}$. The deterministic  attenuation is $1.9\,\textrm{dB}$.
 The output signal is squeezed by $-0.95\,\textrm{dB}$.
 With the postselection protocol, the
 squeezing value can be improved depending on the postselection
threshold $\eta_\textrm{min}$.
 The theory is shown for the elliptic-beam, log-normal, and beam-wandering models,
 compared with the experimental results (shaded area lies within the error bars) from~Ref.\cite{Peuntinger}.
 }
\end{figure}

The postselection procedure of transmission events  with $\eta\ge
\eta_\textrm{min}$ yields larger detected values for the transmitted
squeezing. In Fig.~\ref{fig:sq},
we compare
the values of detected squeezing as functions of postselection thresholds
$\eta_\textrm{min}$, for the experimental values given
in Ref.~\cite{Peuntinger}.
The beam-wandering model  yields smaller values of
postselected
squeezing as detected in the experiment, as it does not properly describe the
distribution tails for high values of $\eta$; cf.
Fig.~\ref{fig:PDTC}. The postselected values of squeezing calculated within the
elliptic-beam approximation   agree very well within the error bars with
the experimentally measured values.
The shown log-normal model  gives the correct values for the first two moments of $\eta$, but it differs in
higher moments from the experimentally measured distribution. This feature allows one to
obtain the correct value of squeezing for the transmitted signal
from the log-normal model. However, this model completely fails to describe the postselection
procedure, where the higher moments play a dominant role.

\paragraph*{Summary and Conclusions.--} We
have introduced a model for the  atmospheric turbulence effects on quantum light,
which is based on an elliptic-beam approximation. Surprisingly, it yields
a reasonable agreement with experiments
for the conditions of weak-to-moderate turbulence. In this case, we
get an excellent description of the transfer of squeezed light through a 1.6 km
channel, analyzed with data postselection.

For the case of strong turbulence, we have shown that our theory gives  a reasonable agreement with the log-normal distribution.  In experiments
using a 144 km channel under strong turbulence conditions, the log-normal model also yields a proper description of the transmission of coherent light. Hence, our theory describes in a unified manner the  quantum-light transfer through atmospheric channels under dissimilar turbulence conditions.
The case of the transition regime of moderate-to-strong turbulence requires further research.

The authors are grateful to P.~Villoresi,
G.~Vallone, Ch.~Marquardt, B.~Heim, and C.~Peuntinger for useful and enlightening
discussions.
The work was supported by the Deutsche Forschungsgemeinschaft through
Project No. VO 501/21-1 and SFB 652, Project No. B12.

\clearpage
\newpage

\onecolumngrid

	\section*{Supplemental Material\\
		Atmospheric Quantum Channels with Weak and Strong Turbulence }

		\begin{center}
		 \textbf{D. Vasylyev$^{1,2}$, A. A. Semenov$^{1,3}$, and W. Vogel$^1$}\\
		$^{1}$\textit{Institut f\"ur Physik, Universit\"at Rostock,
Albert-Einstein-Stra\ss{}e 23, D-18059 Rostock, Germany}\\
		$^{2}$\textit{Bogolyubov Institute for Theoretical Physics, NAS of Ukraine, Vulytsya
Metrologichna 14-b, 03680 Kiev, Ukraine}\\
		$^{3}$\textit{Institute of Physics, NAS of Ukraine, Prospect Nauky 46, 03028
Kiev, Ukraine}
		\end{center}

\noindent The supplement is structured as follows:\\
In Sec.~\ref{Sec:EllipticBeam} we discuss the properties of Gaussian elliptical beams. 
In 
Sec.~\ref{Sec:Transmittance} we derive the analytic expression for the transmittance 
of 
the elliptical beam through the circular aperture. In 
Sec.~\ref{Sec:GasussianApproximation} the statistical properties of the elliptical beam 
transmitted through turbulence  are discussed in Gaussian approximation. In 
Sec.~\ref{Sec:Isotropy} we discuss the simplifications which arise from the assumption 
that the atmospheric turbulence is isotropic. Here we derive the  formulas that connect 
the statistical characteristics of the elliptical beam in the isotropic atmosphere with the 
field correlation functions.  In Sec.~\ref{Sec:PhaseAppr} the  phase approximation of 
the 
Huygens-Kirchhoff method is presented and the general expressions for field 
correlation functions are derived. In Sec.~\ref{Sec:BW} and in 
Sec.~\ref{Sec:BeamShape} we derive the means and (co)variances connected with 
beam wandering and beam shape deformation, respectively.  These results are 
evaluated for limits of weak and strong turbulence and are summarized in the table in  
Sec.~\ref{Sec:CovMatrixEl}.  Finally, in Sec.~\ref{Sec:LogNormal} the log-normal 
distribution for the beam transmittance is considered.

\makeatletter 
\@addtoreset{figure}{section} 
\@addtoreset{equation}{section}
\def\p@subsection{}
\makeatother
\renewcommand{\thesection}{\Alph{section}}
\renewcommand{\thesubsection}{\thesection.\arabic{subsection}}
\renewcommand{\thesubsubsection}{\thesubsection.\arabic{subsubsection}}
\renewcommand{\thefigure}{\thesection\arabic{figure}}
\renewcommand{\theequation}{\thesection\arabic{equation}}

\vspace{5ex}

\twocolumngrid

\section{Elliptic beams}
\label{Sec:EllipticBeam}

In this Section we discuss the properties of elliptical beams, which are
crucial for the consideration of light transferring through the turbulent
atmosphere. In the paraxial approximation the beam amplitude $u(\mathbf{r},z)$
satisfies the equation, cf.~Ref.~\cite{Fante1},
\begin{align}
 2ik\frac{\partial u(\mathbf{r},z)}{\partial z}+\Delta_\mathbf{r}
u(\mathbf{r},z)+2k^2\delta n(\mathbf{r},z) u(\mathbf{r},z)=0,\label{waveEq}
\end{align}
where $k$ is the wave number, $\delta n(\mathbf{r},z)$ is a small fluctuating
part of the index of air refraction, $\mathbf{r}{=}\left(x\,\,y\right)^T$ is
the vector of transverse coordinates.
The boundary condition in the transmitter plane $z=0$ for the initially
Gaussian beam is given by
\begin{align}
 u(\mathbf{r},z{=}0){=}u_0(\mathbf{r}){=}\sqrt{\frac{2}{\pi
W_0^2}}\exp\Bigl[-\frac{1}{W_0^2}|\mathbf{r}|^2{-}\frac{ik
}{2F}|\mathbf{r}|^2\Bigr].\label{Eq:BoundaryConditions}
\end{align}
Here $W_0$ is the beam spot radius, $F$ is the  wavefront radius in the center
of the transmitting aperture at $z{=}0$. The intensity of light is defined as
\begin{equation}
I(\mathbf{r},z)=\left|u(\mathbf{r},z)\right|^2.\label{Eq:IntensityDef}
 \end{equation}
This function can be chosen in the normalized form
\begin{equation}
 \int_{\mathbb{R}^2}\D^2\mathbf{r}\, I(\mathbf{r},z)=1,
\end{equation}
and Eq.~(\ref{waveEq}) implies that this norm preserves for any $z$.
For our purposes it is also important that $I(\mathbf{r},z){\geq}0$.

Consider the transverse Fourier transform of  intensity,
\begin{equation}
 C(\mathbf{k},z)=\int_{\mathbb{R}^2}\D^2\mathbf{r}\,
I(\mathbf{r},z)e^{i\mathbf{k}\cdot\mathbf{r}},\label{Eq:IntSpektrum}
\end{equation}
where $\mathbf{k}{\cdot}\mathbf{r}$ denotes the scalar product of two vectors.
Similar to the cumulative expansion in the probability theory one writes
\begin{equation}
 \ln
C(\mathbf{k},z)=i\mathbf{k}{\cdot}\mathbf{r}_0{-}\frac{1}{8}\mathbf{k}^\mathrm
{ T } \mathbf{S}\mathbf{k} + \ldots, \label{Eq:CumExpInt}
\end{equation}
where
\begin{equation}
 \mathbf{r}_0=\int_{\mathbb{R}^2}\D^2\mathbf{r}\,\mathbf{r}\,
I(\mathbf{r},z)\label{Eq:BeamCentroid}
\end{equation}
is the beam-centroid position,
\begin{eqnarray}
 \mathbf{S}&=&\left( \begin{array}{cc}
S_{xx} & S_{xy}  \\
S_{xy} & S_{yy} \end{array} \right)\label{Eq:MatrixS_Definition}\\
&=&4\int_{\mathbb{R}^2}\D^2\mathbf{r}
\,\left[(\mathbf{r}-\mathbf{r}_0)(\mathbf{r}-\mathbf{r}_0)^\mathrm{T}
\right ] \,
I(\mathbf{r},z)\nonumber
\end{eqnarray}
is the spot-shape matrix. Within the elliptic-beam approximation we suppose that the
expansion~(\ref{Eq:CumExpInt}) in the aperture plane can be restricted by the
second (Gaussian) term. Substituting it into the inversion of
Eq.~(\ref{Eq:IntSpektrum}),
\begin{equation} 
I(\mathbf{r},z)=\frac{1}{\left(2\pi\right)^2}\int_{\mathbb{R}^2}\D^2\mathbf{r}\,
C(\mathbf{k},z)
e^{-i\mathbf{k}{\cdot}\mathbf{r}}.
\label{Eq:IntSpektrumInversion}
\end{equation}
one gets for the intensity of the elliptic beam
\begin{align}
 I(\mathbf{r},z)=\frac{2}{\pi\sqrt{\det\mathbf{S}}}\exp\Bigl[-2({\mathbf{r}}{-}{
\mathbf{r}}_0)^{\rm
T}{\mathbf{S}}^{-1}({\bf r}{-}{\bf r}_0)\Bigr].\label{Eq:Intens}
\end{align}
In the particular case, when the spot-shape matrix is proportional to the
identity matrix, this expression is reduced to the intensity of a circular
Gaussian beam.

\begin{figure}[ht!]
	\includegraphics{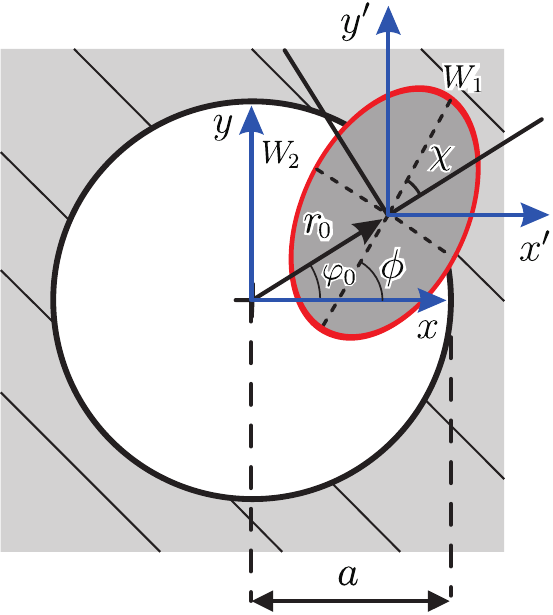}
	\caption{\label{fig:Aperture} (Color online)
		The aperture of radius $a$ and the elliptical beam profile
		with the half-axis $W_{1}$ rotated on the angle $\phi$ relative to the
		$x$-axis and on the angle $\chi$ relative to the $\mathbf{r}_0$-associated
		axis are shown. The
		beam centroid is situated in the point	$\mathbf{r}_0$ with the polar coordinates
		$(r_0,\varphi_0)$. The $x^\prime$-$y^\prime$ coordinate system is associated 
		with the elliptical beam centroid.}
\end{figure}

Two eigenvalues, $W_1^2$ and $W_2^2$, of the spot-shape matrix $\mathbf{S}$
correspond to two semi-axes of the beam ellipse. They are related to the
elements of the matrix $\mathbf{S}$ as
\begin{align}
&S_{xx}=W_1^2\cos^2\!\phi+W_2^2\sin^2\!\phi,\label{Eq:Sxx}\\
 &S_{yy}=W_1^2\sin^2\!\phi+W_2^2\cos^2\!\phi,\label{Eq:Syy}\\
  &S_{xy}=\frac{1}{2}\Bigl(W_1^2-W_2^2\Bigr)\sin2\phi,\label{WxWyrelation}
\end{align}
where
$\phi{\in}\left[0,\pi/2\right)$ is the angle between the $x$-axis
and the ellipse semi-axis related to $W_1^2$. The set of three parameters
$\left(W_1^2,W_2^2,\phi\right)$ uniquely defines all possible orientations of
the ellipse.

The introduced representation of the ellipse assumes that we do not distinguish
between major and minor semi-axes of the ellipse.
The semi-axis related to $W_1^2$ is defined as being situated in first and third
quarter-planes of the $x^\prime$-$y^\prime$ coordinate system, cf.~Fig.~\ref{fig:Aperture}, while 
$W_2^2$ is in the second and fourth ones.
Within this definition the values of $W_1^2$ and  $W_2^2$ are not ordered.

%
%

\section{Aperture transmittance for elliptic beams}
\label{Sec:Transmittance}

In this Section we derive in details an analytical approximation for the transmittance of
elliptic beams through a circular aperture. For the aperture situated in the
point $z{=}L$ the transmittance is determined via the expression,
cf.~Ref.~\cite{Vasylyev2012},
\begin{equation}
 \eta=\int_\mathcal{A}\D^2 \mathbf{r}\, I(\mathbf{r},L),\label{Eq:TransmittanceGeneral}
\end{equation}
where $I(\mathbf{r},L)$ is the normalized intensity defined by
Eq.~(\ref{Eq:IntensityDef}) and integration is performed in the aperture
opening area. Substituting Eq.~(\ref{Eq:Intens}) into
Eq.~(\ref{Eq:TransmittanceGeneral}) and considering the structure of
spot-shape matrix $\mathbf{S}$,
cf.~Eqs.~(\ref{Eq:MatrixS_Definition})~and~(\ref{Eq:Sxx})-(\ref{WxWyrelation}), one gets for
the transmittance,
\begin{align}
\eta&=\frac{2}{\pi W_1 W_2}\int\limits_0^a\D r\,r\int\limits_0^{2\pi}\D\varphi\,
e^{-2A_1\bigl(r\cos\varphi-r_0\bigr)^2}\nonumber\\
&\times e^{-2A_2r^2\sin^2\varphi}
e^{-2A_3\bigl(r\cos\varphi-r_0\bigr)r\sin\varphi}.\label{Tegeneral1}
\end{align}
Here $a$ is the aperture radius, $r$, $\varphi$ are polar
coordinates for the vector  $\mathbf{r}$,
\begin{align}
x{=}r\cos\varphi,\label{Eq:x}\\
y{=}r\sin\varphi,\label{Eq:y}
\end{align}
$r_0$, $\varphi_0$ are polar coordinates for the vector $\mathbf{r}_0$,
\begin{align}
 x_0{=}r_0\cos\varphi_0,\label{Eq:x0}\\
 y_0{=}r_0\sin\varphi_0,\label{Eq:y0}
\end{align}
 \begin{align}
 A_1=\Bigl(\frac{\cos^2(\phi-\varphi_0)}{W_1^2}+\frac{\sin^2(\phi-\varphi_0)}{W_2^2}\Bigr),
\end{align}
\begin{align}
 A_2=\Bigl(\frac{\sin^2(\phi-\varphi_0)}{W_1^2}+\frac{\cos^2(\phi-\varphi_0)}{W_2^2}\Bigr),
\end{align}
\begin{align}
 A_3=\Bigl(\frac{1}{W_1^2}-\frac{1}{W_2^2}\Bigr)\sin2(\phi-\varphi_0),
\end{align}
and $\phi$ is defined with the modulo $\pi/2$ such that $\eta$ in Eq.~(\ref{Tegeneral1}) is
a $\pi/2$-periodical function of $\phi$.

For the given angle $\chi=\phi-\varphi_0$ the transmittance $\eta$ as a function of $r_0$
has a behavior
similar to the transmittance of the circular Gaussian beam with a certain effective
spot-radius $W_\textrm{eff}\!\left(\chi\right)$. Applying the method developed in
Ref.~\cite{Vasylyev2012} one can
write the corresponding approximation,
\begin{align}
\eta=\eta_{0} \exp\left\{-\left[\frac{r_0/a}
{R\left(\frac{2}{W_{\rm
			eff}\left(\phi-\varphi_0\right)}\right)}\right]^{\lambda\bigl(\frac{2}{W_{\rm
			eff}\left(\phi-\varphi_0\right)}\bigr)}\right\}.\label{App:Tapprox}
\end{align}
Here $\eta_0$ is the beam transmittance at $r_0{=}0$, and
\begin{align}
R\left(\xi\right)=\Bigl[\ln\Bigl(2\frac{1-\exp[-\frac{1}{2} a^2
	\xi^2]}{1-\exp[-a^2\xi^2]\I_0\bigl(a^2\xi^2\bigr)}\Bigr)\Bigr]^{-\frac{1}{
		\lambda(\xi)}},\label{Eq:R}
\end{align}
\begin{align}
 \lambda\left(\xi\right)&=2a^2\xi^2\frac{e^{-a^2\xi^2}\I_1(a^2\xi^2)}{1-\exp[
 	-a^2\xi^2 ] \I_0\bigl(a^2\xi^2\bigr)}\nonumber\\
 &{\times}\Bigl[\ln\Bigl(2\frac{1-\exp[-\frac{1}{2} a^2
 \xi^2]}{1-\exp[-a^2\xi^2]\I_0\bigl(a^2\xi^2\bigr)}\Bigr)\Bigr]^{-1}\label{Eq:lambda}
\end{align}
are scale and shape functions, respectively.

The transmittance $\eta_0$ is obtained from Eq.~(\ref{Tegeneral1}) by setting the
beam-centroid position $r_0=0$,
\begin{align}
\eta_0&=\frac{2}{\pi W_1 W_2}\int\limits_0^a \D r\, r\int\limits_0^{2\pi}\D\varphi
\,e^{-\bigl\{\frac{1}{W_1^2}+\frac{1}{W_2^2}\bigr\}r^2}\nonumber\\
&\qquad\times e^
{-\left|\frac{1}{W_1^2}-\frac{1}{W_2^2}\right|r^2\cos2(\varphi-\widetilde\varphi)}\nonumber\\
&=\frac{2}{|W_1W_2|}\int\limits_0^{a^2}\D t\, e^{-\bigl\{\frac{1}{W_1^2}+\frac{1}{W_2^2}\bigr\}t}\,
\I_0\Bigl(\left|\frac{1}{W_1^2}{-}\frac{1}{W_2^2}\right|t\Bigr),
\end{align}
where $\widetilde\varphi{=}\frac{1}{2}\arctan[A_3/(A_1{-}A_2)]$.
It is expressed in terms of the incomplete Lipshitz-Hankel
integral, cf.~Ref.~\cite{Agrest}, as
\begin{align}
	\I_{e_0}(a,z)=\int_0^z \D t e^{-a t}\I_0(t),
\end{align}
that results in
\begin{align}
\eta_0=\frac{2 W_1
W_2}{|W_1^2-W_2^2|}\I_{e_0}\Bigl(\frac{W_1^2+W_2^2}{|W_1^2-W_2^2|},a^2
\frac{|W_1^2-W_2^2|}{W_1^2W_2^2}\Bigr).\label{T0eIe0}
\end{align}
The incomplete Lipshitz-Hankel integral can be evaluated numerically. However, using
the relation between the incomplete Lipshitz-Hankel ($\I_{e_0}$) and Weber ($\widetilde
\Q_0$) integrals \cite{Agrest}, we
can rewrite Eq.~(\ref{T0eIe0}) as
\begin{align}
&\eta_0=1-e^{-a^2\frac{W_1^2+W_2^2}{W_1^2W_2^2}}
\Bigl[\I_0\Bigl(a^2\frac{|W_1^2-W_2^2|}{W_1^2W_2^2}\Bigr)\nonumber\\
&+2\widetilde
\Q_0\Bigl(a^2\frac{(W_1+W_2)^2}{2W_1^2W_2^2},a^2
\frac{|W_1^2-W_2^2|}{W_1^2W_2^2}\Bigr)\Bigr].\label{T0eQ0}
\end{align}
In Ref.~\cite{Vasylyev2012},  an analytical approximation for $\widetilde\Q_0$ is derived.  Applying
here  the same procedure for the approximation of the incomplete Weber integral in
Eq.~(\ref{T0eQ0}) one obtains
\begin{align}
\eta_0&{=}1{-}\I_0\Bigl(a^2\frac{W_1^2{-}W_2^2}{W_1^2W_2^2}\Bigr)e^{-a^2\frac{W_1^2{+}W_2^2}{W_1^2W_2^2}}\nonumber\\
&{-}2\left[1{-}e^{-\frac{a^2}{2}\left(\!\frac{1}{W_1}{-}\frac{1}{W_2}\right)}\!\right]\nonumber\\
&\qquad\times\exp\Biggl[\!{-}\Biggl\{\!\frac{\frac{(W_1+W_2)^2}{|W_1^2-W_2^2|}}{R\left(\frac{1}{W_1}{-}\frac{1}{W_2}\right)}\!\Biggr\}
^{\lambda\left(\!\frac{1}{W_1}{-}\frac{1}{W_2}\right)}\Biggr],\label{T0approx}
\end{align}
where $R\!\left(\xi\right)$ and $\lambda\!\left(\xi\right)$ are defined by Eqs.~(\ref{Eq:R}) and
(\ref{Eq:lambda}), respectively. For the case when $W_1^2{=}W_2^2{=}W^2$,
Eqs.~(\ref{T0eQ0}) and (\ref{T0approx}) are reduced to $\eta_0{=}1{-}e^{-2a^2/W^2}$ that
is the maximal transmittance of the circular beam, cf.~Ref.~\cite{Vasylyev2012}.

In order to  get an approximate value for the effective spot-radius $W_{\rm
eff}\!\left(\chi\right)$ we assume that the intensity of the corresponding circular beam is
equal to the intensity of the elliptic beam at the aperture plane, i.e.
\begin{align}
&\frac{1}{W_\textrm{eff}^2\!\left(\chi\right)} e^{-\frac{2}{W_{\rm
eff}^2\!\left(\chi\right)}(r^2+r_0^2+2\,r\, r_0 \cos\varphi)}=\frac{1}{W_1W_2}\nonumber\\
&\times e^{-2 A_1(\chi) r_0^2} e^{2r_0 r \bigl\{2A_1 (\chi)
\cos\varphi+A_3(\chi)\sin\varphi\bigr\}}\nonumber\\
&\quad\times e^{-2r^2\bigl\{A_2(\chi)+\left[A_1(\chi)-A_2(\chi)\right]\cos^2\varphi+
	\frac{A_3(\chi)}{2}\sin 2\varphi\bigr\}}.\label{Weffdeterm}
\end{align}
In the most general case this equality cannot be satisfied exactly. However, we can
find such a value of  $W_\textrm{eff}\!\left(\chi\right)$ that Eq.~(\ref{Weffdeterm}) will be
fulfilled approximately. For this purpose we expand both sides of this equation in series
with respect to $e^{i\varphi}$. Then we equate the zeroth-order terms of these
expansions at the point $r{=}r_0{=}a$. This results in the expression
\begin{align}
&4\frac{a^2}{W_{\rm eff}^2(\chi)}+\ln\Bigl[\frac{W_{\rm
eff}^2(\chi)}{a^2}\Bigr]-2a^2\left[\frac{1}{W_1^2}+\frac{1}{W_2^2}\right]\nonumber\\
&{-}a^2\left[\frac{1}{W_1^2}{+}\frac{1}{W_2^2}\right]\cos2\chi{-}\ln
\left(\frac{W_1W_2}{a^2}\right)=0.
\end{align}
Solving this equation with respect to $W_\textrm{eff}\left(\chi\right)$ one gets
\begin{align}
W_\textrm{eff}^2\left(\chi\right)&{=}4a^2\Bigl[\mathcal{W}\Bigl(\frac{4a^2}{
	W_1W_2 } e^{\frac{a^2}{W_1^2}\bigl\{1+2\cos^2\!\chi\bigr\}}\nonumber\\
&\qquad\qquad\times
e^
{\frac{a^2}{W_2^2}\bigl\{1+2\sin^2\!\chi\bigr\}}\Bigl)\Bigr]^{-1},\label{App:Weff}
\end{align}
where $\mathcal{W}(x)$ is the Lambert function \cite{Corless}.

\begin{figure}[ht!]
	\includegraphics{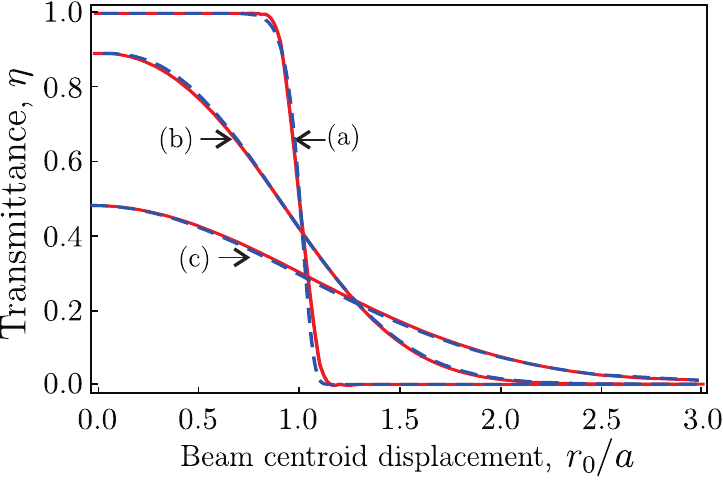}
	\caption{\label{fig:T} (Color online) The transmittance of the elliptical beam (half-axes
		$|W_1|$, $|W_2|$) through the circular aperture (radius $a$) as a function of the beam-centroid
		displacement $r_0$:  (a)~$|W_1|=0.2a,
		|W_2|{=}0.1a$, $\chi={\pi}/{3}$; (b)~$|W_1|{=}a, |W_2|{=}0.9a$, $\chi={\pi}/{4}$;
		(c)~$|W_1|{=}1.8a, |W_2|{=}1.7a$, $\chi={\pi}/{5}$.
		The solid line for $\eta$ is obtained by numerical calculation, the dashed line
		represents the analytical approximation, cf.~Eq.~(\ref{App:Tapprox}).}
\end{figure}

In Fig.~\ref{fig:T} we compare the transmittance $\eta$ obtained by numerical integration
of Eq.~(\ref{Tegeneral1}) and its analytical approximation. The
approximation, cf.~Eq.~(\ref{App:Tapprox}), gives a reasonable accuracy especially in the case
of small beam ellipticity. It is also important to note that
$W_\textrm{eff}^2\left(\phi-\varphi_0\right)$
and $\eta$, cf.~Eqs.~(\ref{App:Weff}) and (\ref{T0approx}), respectively, are $\pi/2$-periodical
functions of the $\phi$, since this angle is defined with the modulo $\pi/2$.

\section{Gaussian approximation}
\label{Sec:GasussianApproximation}

In this Section we discuss in detail the  statistical properties of elliptic beams and discuss the
applicability of the Gaussian approximation. Any spot in the elliptic-beam approximation at
the aperture plane is uniquely described by the set of five parameters
$(x_0,y_0,W_1^2,W_2^2,\phi)$. While the beam passes through the turbulent atmosphere,
these parameters are randomly changed. Each part of the path slightly contributes in
these values. Also it is important to note that these parameters can be correlated.

Random fluctuations of the beam-centroid position $\mathbf{r}_0$, i.e. the parameters
$x_0$ and $y_0$, lead to the effect of beam wandering. These parameters can be
considered as affected by an additive noise during the propagation. A large number of
small additive contributions is a good argument for using the Gaussian approximation for
the beam-centroid position, cf.~Ref.~\cite{Vasylyev2012}.

\subsection{Wrapped Gaussian model for $\phi$}

Similar argumentations work for the angle $\phi$. This parameter can also be considered
as affected by a large number of the small additive contributions. An important difference
is that the angle $\phi$ is a $\pi/2$-periodical variable. For this reason one should use in
this case the wrapped Gaussian distribution, cf.~Ref.~\cite{Mardia},
\begin{equation}
\rho\!\left(\phi\right)=\frac{1}{\sqrt{2\pi}\,\sigma_\phi}\sum\limits_{k=-\infty}^{+\infty}
\exp\left[-\frac{\left(\phi-\mu_\phi+\frac{\pi}{2}k\right)^2}{2\sigma_\phi^2}\right],
\label{Eq:WrappedGaussian}
\end{equation}
where $\mu_\phi$ is the mean direction and $\sigma_\phi$ is the unwrapped standard
deviation. For $\sigma_\phi{\rightarrow}+\infty$
Eq.~(\ref{Eq:WrappedGaussian}) becomes the probability density of the uniform
distribution.

\subsection{Multiplicative-noise model for $W_{i}^2$}

In this model one assumes that each small $k\textrm{th}$ part of the atmospheric channel
multiplicatively changes values of $W_{i}^2$, $i{=}1,2$, with the factor $\varepsilon_{i}^{\,
k}\in\mathbb{R}^{+}$. As a
result at
the aperture plane the value of $W_{i}^2$ is
\begin{align}
W_{i}^2=W_0^2\prod\limits_{k=1}^N\varepsilon_{i}^{\, k},\qquad i{=}1,2.\label{Eq:MultiplicativeNoise}
\end{align}
The large number $N$ of small random
contributions gives a good argument for assuming $W_{i}^2$ log-normally distributed.

Let us introduce the random parameters
\begin{align}
	\Theta_{i}=\ln\frac{W_{i}^2}{W_0^2}.\label{Eq:Theta}
\end{align}
In framework of the considered model these parameters yield a two-fold normal
distribution. For the complete characterization of this distribution we need the means and
the (co)variances of $\Theta_{i}$. They can be expressed in terms of the means and the
(co)variances of $W_{i}^2$ as
\begin{align}
	\langle
	\Theta_{i}\rangle=\ln\left[\frac{\langle
	W_{i}^2\rangle}{W_0^2\left(1+
	\frac{\langle (\Delta	W_{i}^2)^2\rangle}{\langle	
	W_{i}^2\rangle^2}\right)^{1/2}}\right],\label{App:Eq:ThetaMean}
\end{align}
\begin{align}
	\langle \Delta\Theta_i\Delta\Theta_j\rangle=
	\ln\left(1+
	\frac{\langle \Delta W_i^2 \Delta W_j^2\rangle}{\langle	
		W_i^2\rangle\langle	
		W_j^2\rangle}\right),\quad i,j=1,2\label{App:Eq:ThetaCovariances}
\end{align}
which can be used for the corresponding calculations. 

\section{Isotropy of turbulence}
\label{Sec:Isotropy}

In this Section we discuss simplifications, which follow from the assumption that  the
atmospheric turbulence is isotropic. We also assume that
\begin{align}
 \langle\mathbf{r}_0\rangle{=}0,\label{Eq:ZeroBeamCentroid}
\end{align}
i.e. beam wandering fluctuations are placed
around the
reference-frame origin. We consider the field intensity at the aperture plane,
$I(\mathbf{r},L)$, as a stochastic field characterized by the probability
density functional $\rho\left[I(\mathbf{r},L)\right]$. The above assumptions
mean that
\begin{align}
 \rho\left[I(O\,\mathbf{r},L)\right]=\rho\left[I(\mathbf{r},L)\right],
 \label{Eq:Isotropy}
\end{align}
where $O$ is a representation of the  $O(2)$ group. In the following we consider
important consequences from Eqs.~(\ref{Eq:ZeroBeamCentroid}) and
(\ref{Eq:Isotropy}).

\subsection{Uniform distribution for the angle $\phi$}

A clear consequence from the isotropy assumption is the fact that the angle
parameter $\phi$ appears to be uniformly distributed.  This fact is a consequence from
Eq.~(\ref{Eq:Isotropy}). Indeed, according to this requirement the probability density
$\rho(\phi)$ does not depend on the choice of the reference frame, i.e. for any angle $\zeta$
\begin{align}
	\rho(\phi+\zeta)=\rho(\phi).
\end{align}
This equation holds true only for the uniform distribution.
%
For details of circular distributions see Ref.~\cite{Mardia}.

\subsection{Correlations between linear and angle parameters}

Let $\mathbf{v}$ be a random vector, which consists of variables $v_i$ with the support
$\mathbb{R}$,
\begin{align}
\mathbf{v}=\left(\begin{array}{cccc}
x_0&y_0&\Theta_1&\Theta_2
\end{array}\right)^\mathrm{T}.\label{Eq:vMultNoise}
\end{align}
%
The parameters $v_i,\, i{=}1,..,4$ of Eq.~(\ref{Eq:vMultNoise}) and the angle parameter $\phi$ are 
distributed according to the two-fold normal
distribution, which is wrapped for $\phi$, cf.~Section~\ref{Sec:GasussianApproximation},
\begin{align}
	\rho\left(v_i,\phi\right)=\frac{1}{2\pi\sqrt{\det\Sigma_{v_i,\phi}}}
	\sum\limits_{k=-\infty}^{+\infty}
	\exp\left(-\frac{1}{2}\boldsymbol{\nu}_k^\mathrm{T}\,\Sigma_{v_i,\phi}^{-1}\,
	\boldsymbol{\nu}_k\right),
	\label{Eq:TwoFoldWrapped}
\end{align}
where
\begin{align}
\boldsymbol{\nu}_k=\left(\begin{array}{cc}
v_i-\langle v_i\rangle&\phi-\mu_\phi+\frac{\pi}{2}k
\end{array}\right)^\mathrm{T},
\end{align}
$\mu_\phi$ is the mean direction of $\phi$
\begin{align}
	\Sigma_{v_i,\phi}=\left(\begin{array}{cc}
	\sigma_{v_i}^2&s\sigma_{v_i}\sigma_\phi\\
	s\sigma_{v_i}\sigma_\phi&\sigma_\phi^2
	\end{array}\right)
\end{align}
is the covariance matrix, $\sigma_{v_i}^2$ is the standard deviation of $v_i$,
$\sigma_\phi^2$ is the unwrapped variance of $\phi$, and $s$ is the correlation
coefficient.

The considered probability distribution can also be rewritten in the form, cf.~Ref.~\cite{Mardia},
\begin{align}
\rho&\left(v_i,\phi\right)=
\frac{1}{\sqrt{2\pi}\sigma_{v_i}}e^{-\frac{1}{2}\frac{(v_i-\langle v_i
\rangle)^2}{\sigma_{v_i}^2}}\frac{2}{\pi}
\Bigl\{1\label{wrapped}\\
&{+}2\sum\limits_{n=1}^\infty
e^{-8(1{-}s^2)\sigma_\phi^2n^2}\!
\cos\Bigl[4n\bigl(\phi{-}\mu_\phi{-}s\frac{\sigma_\phi}{\sigma_{v_i}}[v_i{-}\langle
 v_i\rangle]\bigr)\Bigr]\Bigr\}.\nonumber
\end{align}
As it has been already shown, in the case of isotropic turbulence the marginal distribution
for $\phi$ is uniform. This corresponds to the case of $\sigma_\phi^2{\rightarrow}+\infty$.
If the correlation is imperfect, i.e. $s^2{\neq}1$, Eq.~(\ref{wrapped}) is factorized in the
normal distribution for $v_i$ and the uniform distribution for $\phi$,
\begin{align}
\rho\left(v_i,\phi\right)=
\frac{1}{\sqrt{2\pi}\sigma_{v_i}}e^{-\frac{1}{2}\frac{(v_i-\langle v_i
		\rangle)^2}{\sigma_{v_i}^2}}\frac{2}{\pi},\quad i{=}1,...,4.\label{Eq:GaussUniform}
\end{align}
Hence, for the isotropic turbulence correlations between the angle $\phi$ and the
linear parameters vanish.

\subsection{Correlations between beam-centroid position and spot-shape
	parameters}
\label{Sec:CorrR0W}

Consider the random variables $\Theta_{i}$, $i{=}1,2$, which describe the spot shape, 
cf.~Eq.~(\ref{Eq:Theta}). We will be interested in the correlations $\langle \Delta
\Theta_{i}\,\Delta\mathbf{r}_0\rangle$. With the considered assumption
\begin{align}
	\langle \Delta \Theta_{i}\,\Delta \mathbf{r}_0\rangle=\langle 
	\Theta_{i}\,\mathbf{r}_0\rangle,\qquad i{=}1,2,
\end{align}
because the beam centroid is fluctuating around the reference-frame origin,
cf.~Eq.~(\ref{Eq:ZeroBeamCentroid}).

By using the definition of $\mathbf{r}_0$, cf.~Eq.~(\ref{Eq:BeamCentroid}), the correlation
coefficient is written as
\begin{align}
	\langle \Delta \Theta_{i}\,\Delta\mathbf{r}_0\rangle
	=\int_{\mathbb{R}^2}\D^2\mathbf{r}\,\mathbf{r}\,
	\langle\Theta_{i}\, I(\mathbf{r},L)\rangle.\label{Eq:CorrXiX0}
\end{align}
The assumption of isotropy, cf.~Eq.~(\ref{Eq:Isotropy}), results in the statement that
$\langle\Theta_{i}\, I(\mathbf{r},L)\rangle$ is invariant with respect to the rotations in the 
$(x,y)$
plane.
Hence, this function has the central symmetry. This leads to the conclusion that
\begin{align}\label{Thetacor}
	\langle \Delta\Theta_{i}\,\Delta\mathbf{r}_0\rangle
	=0,
\end{align}
because the integral in Eq.~(\ref{Eq:CorrXiX0}) vanishes.

We assume that $\Theta_{i}$ and $\mathbf{r}_0$ are Gaussian variables,
cf.~Section~\ref{Sec:GasussianApproximation}. Together with Eq.~(\ref{Thetacor}) 
this yields
\begin{align}
	\langle F(\Theta_{i})G(\mathbf{r}_0)\rangle=\langle F(\Theta_{i})\rangle\,\langle
	G(\mathbf{r}_0)\rangle.
\end{align}
Here $F$ and $G$ are arbitrary functions.

\subsection{Moments and (co)variances of $W_{i}^2$}

In Section~\ref{Sec:GasussianApproximation} it has been shown that for the
characterization of probability distributions for elliptic beams we need among other first
and second moments for $W_{i}^2$, $i{=}1,2$, cf.~Eqs.~(\ref{App:Eq:ThetaMean}) and
(\ref{App:Eq:ThetaCovariances}). In general, the calculation
of these moments is a complicated task, which requires non-Gaussian functional
integration. Here we will show that the assumption of turbulence isotropy essentially
simplifies this problem such that the moments are expressed in terms of field correlation
functions of the second and fourth orders.

\subsubsection{First moments of $W_{i}^2$}
We start the consideration with averaging the elements of the matrix $\mathbf{S}$,
cf.~Eq.~(\ref{Eq:MatrixS_Definition}), by the
atmosphere states,
\begin{align}
	\langle &S_{xx}\rangle{=}\nonumber\\
	&4\left[\int_{\mathbb{R}^2}\D^2\mathbf{r}\, x^2 \Gamma_2\!
	\left(\mathbf{r};L\right){-}
	\int_{\mathbb{R}^4}\D^2\mathbf{r}_1\D^2\mathbf{r}_2\,x_1x_2
	\Gamma_4\!\left(\mathbf{r}_1,\mathbf{r}_2;L\right)\right],\label{Eq:SxxMean}\\
	\langle &S_{yy}\rangle{=}\nonumber\\
	&4\left[\int_{\mathbb{R}^2}\D^2\mathbf{r}\, y^2 \Gamma_2\!
	\left(\mathbf{r};L\right){-}
	\int_{\mathbb{R}^4}\D^2\mathbf{r}_1\D^2\mathbf{r}_2\,y_1y_2
	\Gamma_4\!\left(\mathbf{r}_1,\mathbf{r}_2;L\right)\right],\label{Eq:SyyMean}\\
	\langle &S_{xy}\rangle{=}\nonumber\\
	&4\left[\int_{\mathbb{R}^2}\D^2\mathbf{r}\, xy \Gamma_2\!
	\left(\mathbf{r};L\right){-}
	\int_{\mathbb{R}^4}\D^2\mathbf{r}_1\D^2\mathbf{r}_2\,x_1y_2
	\Gamma_4\!\left(\mathbf{r}_1,\mathbf{r}_2;L\right)\right].\label{Eq:SxyMean}
\end{align}
Here
\begin{align}
	\Gamma_2\!\left(\mathbf{r};z\right)=\left\langle I(\mathbf{r},z)\right\rangle=
	\left\langle u^\ast(\mathbf{r},z)u(\mathbf{r},z)\right\rangle,\label{Eq:Gamma2}
\end{align}
    \begin{align}
	\Gamma_4\!\left(\mathbf{r}_1,\mathbf{r}_2;z\right)&=\left\langle
	I(\mathbf{r}_1,z)I(\mathbf{r}_2,z)\right\rangle\label{Eq:Gamma4}\\&=
	\left\langle u^\ast(\mathbf{r}_1,z)u(\mathbf{r}_1,z)
	u^\ast(\mathbf{r}_2,z)u(\mathbf{r}_2,z)\right\rangle\nonumber
	\end{align}
are the field correlation functions of the second and fourth orders, respectively. The
isotropy assumption, cf.~Eq.~(\ref{Eq:Isotropy}), results in the equalities,
\begin{align}
	&\langle S_{xx}\rangle{=}\langle S_{yy}\rangle,\label{Eq:SxxEqSyy}\\
	&\langle S_{xy}\rangle{=}0,\label{Eq:SxyEq0}
\end{align}
which means that the averaged beam has a circular shape. Equation~(\ref{Eq:SxyEq0}) is
a consequence of the fact that due to the turbulence isotropy
$\Gamma_2\!\left(\mathbf{r};L\right)$ and $\int_{\mathbb{R}^2}\D x_2\D y_1\,x_1y_2
\Gamma_4\!\left(\mathbf{r}_1,\mathbf{r}_2;L\right)$ have a symmetry in
planes $(x,y)$ and $(x_1,y_2)$, respectively. This symmetry implies that the integrals in
Eq.~(\ref{Eq:SxyMean}) appear to have zero values.

Combining Eqs.~(\ref{WxWyrelation}) and (\ref{Eq:SxyEq0}) one gets
\begin{align}
	\left\langle W_1^2\right\rangle=\left\langle W_2^2\right\rangle,
\end{align}
 where we have used the fact that the angle $\phi$ does not correlate with $W_{i}^2$,
 cf. Eq.~(\ref{Eq:GaussUniform}). Similarly, averaging Eqs.~(\ref{Eq:Sxx}) and (\ref{Eq:Syy})
 one gets
 \begin{align}
 	\left\langle W_{1/2}^2 \right\rangle=\left\langle S_{xx/yy} \right\rangle.
 	\label{Eq:W12EqSxxyy}
 \end{align}
 This equation together with Eqs.~(\ref{Eq:SxxMean}) and (\ref{Eq:SyyMean}) express the
 first  moments of $W_{i}^2$ in terms of the field correlation functions $\Gamma_2$ and
 $\Gamma_4$.

 \subsubsection{Second moments of $W_{i}^2$}

 Similar argumentations enable us to express the second moments of $W_{i}^2$, $i{=}1,2$, in terms
 of field correlation functions. For this purpose we multiply Eq.~(\ref{WxWyrelation}) by
 $(W_1^2+W_2^2)$ and average it,
 \begin{align}
 	\left\langle S_{xy}W_1^2\right\rangle+\left\langle S_{xy}W_2^2\right\rangle=
 	\frac{1}{2}\Bigl(\left\langle W_1^4\right\rangle-	
 	\left\langle W_2^4\right\rangle\Bigr)
 		\left\langle\sin2\phi\right\rangle.\label{Eq:W4Derivation1}
 \end{align}
 Here
 \begin{align}
 	\left\langle S_{xy}W_{i}^2\right\rangle&=
 	4\left[\int_{\mathbb{R}^2}\D^2\mathbf{r}\, xy \left\langle
 	W_{i}^2 I(\mathbf{r},L)\right\rangle\right.\label{Eq:W4Derivation2}\\
 	&\left.{-}\int_{\mathbb{R}^4}\D^2\mathbf{r}_1\D^2\mathbf{r}_2\,x_1y_2
 	\left\langle W_{i}^2
 	I(\mathbf{r}_1,L)I(\mathbf{r}_2,L)\right\rangle\right].\nonumber
 \end{align}
The isotropy condition~(\ref{Eq:Isotropy}) implies that the functions $\left\langle
W_{i}^2 I(\mathbf{r},L)\right\rangle$ and $\int_{\mathbb{R}^2}\D x_2\D y_1\,
\left\langle W_{i}^2 I(\mathbf{r}_1,L)I(\mathbf{r}_2,L)\right\rangle$ have such a symmetry
in $(x,y)$ and $(x_1,y_2)$ planes, respectively, that the integrals in
Eq.~(\ref{Eq:W4Derivation2}) are zeros. This means that the left-hand side of
Eq.~(\ref{Eq:W4Derivation1}) is also zero, which results in
\begin{align}
	\left\langle W_1^4\right\rangle=\left\langle W_2^4\right\rangle.\label{Eq:W14EqW24}
\end{align}
The assumption of isotropy also implies that
\begin{align}
	\langle S_{xx}^2\rangle{=}\langle S_{yy}^2\rangle,\label{Eq:Sxx2EqSyy2}
\end{align}
i.e. the second moments of $S_{xx/yy}$ are equal.

Equations~(\ref{Eq:Sxx}) and (\ref{Eq:Syy}) enable to express the moments $\langle
S_{xx/yy}^2\rangle$ and $\langle S_{xx}S_{yy}\rangle$ in terms of the moments
$\langle W_{1/2}^4\rangle$ and $\langle W_{1}^2W_{2}^2\rangle$,
\begin{align}
	&\langle S_{xx/yy}^2\rangle=\frac{3}{4}\langle W_{1/2}^4\rangle
	+\frac{1}{4}\langle W_{1}^2W_{2}^2\rangle,\label{Eq:Sxxyy2Mean}\\
	&\langle S_{xx}S_{yy}\rangle=\frac{1}{4}\langle W_{1/2}^4\rangle+\frac{3}{4}\langle
	W_{1}^2W_{2}^2\rangle,\label{Eq:SxxSyyMean}
\end{align}
where we have utilized the absence of correlations between $W_{1/2}^2$ and the angle
$\phi$, cf.~Eq.~(\ref{Eq:GaussUniform}). Inverting Eqs.~(\ref{Eq:Sxxyy2Mean}) and
(\ref{Eq:SxxSyyMean}) one gets
\begin{align}
&\langle W_{1/2}^4\rangle=\frac{3}{2}\langle S_{xx/yy}^2\rangle
-\frac{1}{2}\langle S_{xx}S_{yy}\rangle,\label{Eq:W12_2Mean}\\
&\langle W_{1}^2W_{2}^2\rangle=-\frac{1}{2}\langle S_{xx/yy}^2\rangle
+\frac{3}{2}\langle S_{xx}S_{yy}\rangle.\label{Eq:W1W2Mean}
\end{align}
Since the moments $\langle S_{xx/yy}^2\rangle$ and $\langle S_{xx}S_{yy}\rangle$
can be expressed in terms of field correlation functions, we get a tool for obtaining the
moments $\langle W_{1/2}^4\rangle$ and $\langle W_{1}^2W_{2}^2\rangle$.

The straightforward expressions for  $\langle S_{xx/yy}^2\rangle$ and $\langle
S_{xx}S_{yy}\rangle$ contain the even-order  field correlation functions up to
$\Gamma_8$. Analytical methods are quite involved for evaluation of
sixth- and eight-order functions. By using the assumptions of Gaussianity for the
beam parameters and isotropic properties of the turbulence we can rewrite these
expressions in terms of field correlation functions $\Gamma_2$ and $\Gamma_4$ only.

The moment $\langle S_{xx}^2\rangle$ is obtained from
Eq.~(\ref{Eq:MatrixS_Definition}) by squaring and averaging $S_{xx}^2$,
\begin{align}
	\langle
	S_{xx}^2\rangle=&16\left(\int_{\mathbb{R}^4}\D^2\mathbf{r}_1\D^2\mathbf{r}_2\,x_1^2x_2^2
	\,\Gamma_4\!\left(\mathbf{r}_1,\mathbf{r}_2;L\right)+\langle
	x_0^4\rangle\right.\label{Eq:SxxViaGamma1}\\
	&\left.-2\left\langle x_0^2\int_{\mathbb{R}^2}\D^2\mathbf{r}\,x^2
	\,I\!\left(\mathbf{r};L\right)\right\rangle\right),\nonumber
\end{align}
and similarly for the moment $\langle S_{yy}^2\rangle$. The second term on the right-hand
side of this expression contains the field correlation function $\Gamma_8$. However,
assuming that the beam-centroid coordinate, $x_0$, is a Gaussian variable and utilizing
Eq.~(\ref{Eq:ZeroBeamCentroid}), this term can be written as
\begin{align}
	\langle x_0^4\rangle=3\langle x_0^2\rangle^2.\label{Eq:FourthMomentX0}
\end{align}
Here
\begin{align}
	\langle x_0^2\rangle=\int_{\mathbb{R}^4} \D^2 \mathbf{r}_1\D^2 \mathbf{r}_2 x_1 x_2\,
	\Gamma_4(\mathbf{r}_1,\mathbf{r}_2;L),\label{App:bwvariance}
\end{align}
which is expressed in terms of the field correlation function $\Gamma_4$.

Consider the third term in right-hand side of Eq.~(\ref{Eq:SxxViaGamma1}). By using
Eqs.~(\ref{Eq:MatrixS_Definition}), (\ref{Eq:Sxx}), (\ref{Eq:GaussUniform}), and
(\ref{Eq:FourthMomentX0}) one gets
\begin{align}
	\left\langle x_0^2\int_{\mathbb{R}^2}\D^2\mathbf{r}\,x^2
	\,I\!\left(\mathbf{r};L\right)\right\rangle=\frac{1}{4}\langle x_0^2 W_{1/2}^2\rangle+3\langle
	x_0^2\rangle^2.
\end{align}
Because the assumption of isotropy results in the fact that the beam-centroid coordinate
$x_0$ does not correlate with the spot-shape parameters, cf.
Section~\ref{Sec:CorrR0W}, we can write
\begin{align}
	\langle x_0^2 W_{1/2}^2\rangle=\langle x_0^2\rangle \langle S_{xx/yy}\rangle,
\end{align}
where we have also used Eq.~(\ref{Eq:W12EqSxxyy}). Next, the expression for the
moment $\langle S_{xx}^2\rangle$, cf.~Eq.~(\ref{Eq:SxxViaGamma1}), in terms of the
second- and fourth-order field correlation functions reads as
\begin{align}
\langle
S_{xx}^2\rangle=&16\left(\int_{\mathbb{R}^4}\D^2\mathbf{r}_1\D^2\mathbf{r}_2\,x_1^2x_2^2
\,\Gamma_4\!\left(\mathbf{r}_1,\mathbf{r}_2;L\right)\right.\label{Eq:SxxViaGamma2}\\
&\left.-3\langle
x_0^2\rangle^2
-\frac{1}{2}\langle x_0^2\rangle \langle S_{xx}\rangle\right).\nonumber
\end{align}
Similar considerations should be applied for the calculation of the moment $\langle
S_{xx}S_{yy}\rangle$, taking into account that $\langle x_0^2y_0^2\rangle=\langle x_0^2\rangle^2$.

Finally, we substitute the obtained expressions for the moments $\langle S_{xx}^2\rangle$
and $\langle S_{xx}S_{yy}\rangle$ in Eqs.~(\ref{Eq:W12_2Mean}) and
(\ref{Eq:W1W2Mean}). This results in relations for the moments $\langle
W_{1/2}^4\rangle$ and $\langle W_{1}^2W_{2}^2\rangle$ in terms of field correlation
functions $\Gamma_2$ and $\Gamma_4$,
\begin{align}
\langle
&W_{1/2}^4\rangle{=}8\left(
3\int_{\mathbb{R}^4}\D^2\mathbf{r}_1\D^2\mathbf{r}_2\,x_1^2x_2^2
\,\Gamma_4\!\left(\mathbf{r}_1,\mathbf{r}_2;L\right)\right.\label{App:Eq:W12ViaGamma}\\
&\left.{-}\int_{\mathbb{R}^4}\D^2\mathbf{r}_1\D^2\mathbf{r}_2\,x_1^2y_2^2
\,\Gamma_4\!\left(\mathbf{r}_1,\mathbf{r}_2;L\right){-}8\langle
x_0^2\rangle^2
{-}\langle x_0^2\rangle \langle S_{xx}\rangle\right),\nonumber
\end{align}
\begin{align}
&\langle
W_{1}^2W_{2}^2\rangle{=}8\left(
3\int_{\mathbb{R}^4}\D^2\mathbf{r}_1\D^2\mathbf{r}_2\,x_1^2y_2^2
\,\Gamma_4\!\left(\mathbf{r}_1,\mathbf{r}_2;L\right)\right.\label{Eq:W1W2ViaGamma}\\
&\qquad\left.{-}\int_{\mathbb{R}^4}\D^2\mathbf{r}_1\D^2\mathbf{r}_2\,x_1^2x_2^2
\,\Gamma_4\!\left(\mathbf{r}_1,\mathbf{r}_2;L\right)
{-}\langle x_0^2\rangle \langle S_{xx}\rangle\right).\nonumber
\end{align}
Here  $\langle S_{xx}\rangle$ and $\langle x_0^2\rangle$ are given by
Eqs.~(\ref{Eq:SxxMean}) and (\ref{App:bwvariance}), respectively.

\section{Phase approximation of the Huygens-Kirchhoff method}
\label{Sec:PhaseAppr}

The parameters, which characterize statistical properties of elliptic beams,
are expressed in terms of the field correlation functions $\Gamma_2$ and
$\Gamma_4$, see Section~\ref{Sec:Isotropy}. Here we briefly discuss the method
of obtaining these functions as proposed in Ref.~\cite{Banakh}. We start from the
paraxial equation, cf.~Eq.~(\ref{waveEq}), which describes the beam amplitude,
$u(\mathbf{r},z)$ and the corresponding boundary condition,
$u_0(\mathbf{r}^\prime)$, cf.~Eq.~(\ref{Eq:BoundaryConditions}).
For our purposes this equation is represented in such an integral
form,
\begin{align}
&u(\mathbf{r},z)=\int_{\mathbb{R}^2} \D^2 \mathbf{r}^\prime
u_0(\mathbf{r}^\prime)
G_0(\mathbf{r},\mathbf{r}^\prime;z,0)\,
G_1(\mathbf{r},\mathbf{r}^\prime;z,0)\nonumber
\\
&{+}\frac{i}{2k} \int\limits_{0}^z\D z^\prime
\int_{\mathbb{R}^2}\! \D^2 \mathbf{r}^\prime
 u(\mathbf{r}^\prime,z^\prime)
 G_0(\mathbf{r},\mathbf{r}^\prime;z,z^\prime)
\Delta^\prime
G_1(\mathbf{r},\mathbf{r}^\prime;z,z^\prime).
\label{Eq:IntegraEqU}
\end{align}
Here
\begin{align}
 G_0(\mathbf{r},\mathbf{r}^\prime;z,z^\prime)=
 \frac{k}{2\pi i
(z-z^\prime)}\exp\Bigl[\frac{ik
|\mathbf{r}-\mathbf{r}^\prime|^2}{2(z-z^\prime)}\Bigr],\label{Eq:G0}
\end{align}
\begin{align}
 G_1(\mathbf{r},\mathbf{r}^\prime;z,z^\prime)=\exp\Bigl[i
S(\mathbf{r},\mathbf{r}^\prime;z,z^\prime)\Bigr],
\end{align}
\begin{align}
 S(\mathbf{r},\mathbf{r}^\prime;z,z^\prime)
 =k\int\limits_{z^\prime}^z\D \xi\, \delta n\Bigl(
\mathbf{r}\frac{\xi-z^\prime}{z-z^\prime}+\mathbf{r}^\prime\frac{z-\xi}{
z-z^\prime},\xi\Bigr),\label{SfunctDef}
\end{align}
and $\Delta^\prime$ is the transverse Laplace operator acting on functions of
$\mathbf{r}^\prime$.

The phase approximation assumes that we consider the zero-order
approximation for the solution of Eq.~(\ref{Eq:IntegraEqU}) in the aperture plane
$z{=}L$, i.e.
\begin{align}
 u(\mathbf{r},L)=\int_{\mathbb{R}^2} \D^2 \mathbf{r}^\prime
u_0(\mathbf{r}^\prime)
G_0\,
G_1(\mathbf{r},\mathbf{r}^\prime;L,0).
\label{Eq:IntegraEqUZeroSolution}
\end{align}
Substituting this expression in the definition of the field correlation
functions, cf.~Eqs.~(\ref{Eq:Gamma2}) and (\ref{Eq:Gamma4}), one gets
\begin{align}	
&\Gamma_{2n}\!\left(\mathbf{r}_1,\ldots,\mathbf{r}_n;L\right)=
\label{Eq:Gamma2nSol1}\\
	&\int_{\mathbb{R}^{4n}}\D^2\mathbf{r}_1^\prime\ldots
\D^2\mathbf{r}_{2n}^\prime\,
u_0(\mathbf{r}_1^\prime)u_0^\ast(\mathbf{r}_2^\prime)\ldots
u_0(\mathbf{r}_{2n-1}^\prime)u_0^\ast(\mathbf{r}_{2n}^\prime)
\nonumber\\
&\hspace{6em}{}\times\mathcal{G}_{2n,0}(\mathbf{r}_1,\ldots,\mathbf{r}_n,
\mathbf {r}_1^\prime, \ldots,\mathbf{r}_{2n}^\prime;L,0)\nonumber\\
&\hspace{6em}{}\times\left\langle\mathcal{G}_{2n,1}(\mathbf{r}_1,\ldots,
\mathbf{r}_n,\mathbf{r}_1^\prime,\ldots,\mathbf{r}_{2n}^\prime;L,0)\right\rangle
,\nonumber
\end{align}
where $n{=}1,2,\ldots$,
\begin{align}
&\mathcal{G}_{2n,i}(\mathbf{r}_1,\ldots,\mathbf{r}_n,
\mathbf {r}_1^\prime, \ldots,\mathbf{r}_{2n}^\prime;L,0)=\\
&\hspace{5em}{}\prod\limits_{k=1}^n
G_i(\mathbf{r}_k,\mathbf{r}_{2k-1}^\prime;L,0)\,
G_i^\ast(\mathbf{r}_k,\mathbf{r}_{2k}^\prime;L,0),\nonumber
\end{align}
and $i{=}0,1$. The assumption that $\delta n(\mathbf{r};z)$ is a Gaussian
stochastic field enables to average $\mathcal{G}_{2n,1}$ in
Eq.~(\ref{Eq:Gamma2nSol1}), such that
\begin{align}
 &\left\langle\mathcal{G}_{2n,1}(\mathbf{r}_1,\ldots,
\mathbf{r}_n,\mathbf{r}_1^\prime,\ldots,\mathbf{r}_{2n}^\prime;L,
0)\right\rangle=\label{Eq:G2n1}\\
&\qquad{}\exp\Bigl[\frac{1}{2}\sum\limits_{k=2}^{2n}\sum\limits_{l=1}^{k-1}
(-1)^{k+l}\mathcal{D}_S(\mathbf{r}_l,\mathbf{r}_k;\mathbf{r}_l^\prime,
\mathbf{r}_k^\prime;L,0)\Bigr]. \nonumber
\end{align}
Here
\begin{align}
 &\mathcal{D}_S(\mathbf{r}_l,\mathbf{r}_k;\mathbf{r}_l^\prime,
\mathbf{r}_k^\prime;L,0)\label{Eq:StrConstD_S}\\
&\qquad{}=\left\langle\Bigl[S(\mathbf{r}_l,\mathbf{r}_l^\prime;L,0)-
S(\mathbf{r}_k,\mathbf{r}_k^\prime;L,0)\Bigr]^2\right\rangle\nonumber
\end{align}
is the structure function of phase fluctuations of a spherical wave propagating
in turbulence.

The correlation function for the index of refraction in the Markovian
approximation, cf.~e.g.~Ref.~\cite{Fante1}, reads as
\begin{align}\label{DeltaNcorr}
& \langle\delta n(\mathbf{r};z)\delta
n(\mathbf{r}
^\prime;z^\prime)\rangle\\
&\qquad{}=2\pi\delta(z-z^\prime)\int_{\mathbb{R}^2}\D^2\boldsymbol{\kappa}\,
\Phi_n(\boldsymbol{\kappa},z)e^{i\boldsymbol{\kappa}\cdot(\mathbf{r}
-\mathbf{r} ^\prime)}.\nonumber
\end{align}
Here $\Phi_n(\boldsymbol{\kappa},z)$ is the spectrum of turbulence, which we
use in the Kolmogorov form, see Ref.~\cite{Tatarskii},
\begin{align}
 \Phi_n(\boldsymbol{\mathbf{\kappa}},z)=0.033 
 C_n^2(z)\kappa^{-\frac{11}{3}},\label{KolmogorovSpectr}
\end{align}
and $C_n^2(z)$ is the refractive index structure constant. Inserting Eqs.~(\ref{SfunctDef}), 
(\ref{DeltaNcorr}) and (\ref{KolmogorovSpectr}) in
Eq.~(\ref{Eq:StrConstD_S}), we arrive at the following expression for the phase
structure function
\begin{align}
 \mathcal{D}_S(\mathbf{r},\mathbf{r}^\prime)
 =2\rho_0^{-\frac{5}{3}}\int\limits_0^1\D\xi\,
\Bigl|\mathbf{r}\,\xi{+}
\mathbf{r}^\prime(1-\xi)
\Bigr|^ { \frac {5}{3}},\label{DsSph}
\end{align}
where we assume that $C_n^2$ is constant for the horizontal link,
\begin{align}
 \mathcal{D}_S(\mathbf{r}_k-\mathbf{r}_l,\mathbf{r}_k^\prime-
\mathbf{r}_l^\prime)=\mathcal{D}_S(\mathbf{r}_l,\mathbf{r}_k;\mathbf{r}
_l^\prime,
\mathbf{r}_k^\prime;L,0),\label{DsSph1}
\end{align}
is a simplified notion for the structure function of phase fluctuations,
\begin{align}\label{rhoDefinition}
\rho_0=(1.5\, C_n^2\,k^2 L)^{-3/5}
\end{align}
is the radius of spatial coherence of a plane wave in the atmosphere.

Finally we substitute  Eqs.~(\ref{DsSph}), (\ref{DsSph1}) into Eq.~(\ref{Eq:G2n1}). Then substituting 
Eqs.~(\ref{Eq:G0}) and (\ref{Eq:G2n1}) into Eq.~(\ref{Eq:Gamma2nSol1}) and performing some trivial 
integrations, we evaluate  the field correlation functions  for $n=1,2$,
\begin{align}
 &\Gamma_2(\mathbf{r})=\frac{k^2}{4\pi^2 L^2}\int_{\mathbb{R}^2}\D^2 \mathbf{r}^\prime
 e^{-\frac{g^2|\mathbf{r}^\prime|^2}{2W_0^2}
 -2i\frac{\Omega}{W_0^2}\mathbf
 r\cdot\mathbf{r}^\prime-\frac{1}{2}\mathcal{D}_S(0,\mathbf{r}^\prime)}\label{Gamma2}
\end{align}
and
\begin{align}
 &\Gamma_4(\mathbf{r}_1,\mathbf{r}_2)=\frac{2 k^4 }{\pi^2(2\pi)^3L^4
 W_0^2}\int_{\mathbb{R}^6}\D^2 \mathbf{r}^\prime_{1}\D^2 \mathbf{r}^\prime_{2}\D^2
 \mathbf{r}^\prime_{3}\nonumber\\
 &\qquad{\times}e^{-\frac{1}{W_0^2}(|\mathbf{r}_1^\prime|^2+|\mathbf{r}_2^\prime|^2+g^2|\mathbf{r}_3^\prime|^2)+2i\frac{\Omega}{W_0^2}[1{-}\frac{L}{F}]\mathbf{r}^\prime_1\cdot
  \mathbf{r}^\prime_2}
 \nonumber\\
 &\qquad\quad\times
 e^{-2i\frac{\Omega}{W_0^2}(\mathbf{r}_1-\mathbf{r}_2)\cdot\mathbf{r}^\prime_2-2i\frac{\Omega}{W_0^2}(\mathbf{r}_1+\mathbf{r}_2)\cdot
  \mathbf{r}^\prime_3}\nonumber\\
 &\quad\times\exp\Biggl[\frac{1}{2}\sum\limits_{j=1,2}\Bigl\{\mathcal{D}_S(\mathbf{r}_1{-}\mathbf{r}_2,\mathbf{r}^\prime_1{+}(-1)^j
  \mathbf{r}^\prime_2)\nonumber\\
 &{-}
 \mathcal{D}_S(\mathbf{r}_1{-}\mathbf{r}_2,\mathbf{r}^\prime_1{+}(-1)^j
 \mathbf{r}^\prime_3){-} \mathcal{D}_S(0,\mathbf{r}^\prime_2{+}(-1)^j
 \mathbf{r}^\prime_3)\Bigr\}\Biggr].\label{Gamma4}
\end{align}
Here 
\begin{align}
\Omega{=}\frac{kW_0^2}{2L}\label{FresnelOmega} 
\end{align}
 is the Fresnel number of  the transmitter aperture and $g^2{=}1{+}\Omega^2[1{-}\frac{L}{F}]^2$ is 
 the generalized diffraction beam parameter.

\section{Beam wandering}
\label{Sec:BW}
In this Section we derive the beam-wandering variance for weak and strong turbulence regimes. The 
beam-wandering variance $\langle x_0^2\rangle$ is evaluated by substituting  Eq.~(\ref{Gamma4}) 
into  Eq.~(\ref{App:bwvariance})
\begin{align}\label{InitialBW}
 &\langle 
 x_0^2\rangle=\frac{2k^4}{\pi^2(2\pi)^3L^4W_0^2}\int_{\mathbb{R}^{10}}\D^2\mathbf{R}\,\D^2\mathbf{r}\,\D^2\mathbf{r}_1^\prime\,\D^2\mathbf{r}_2^\prime\,\D^2\mathbf{r}_3^\prime\nonumber\\
 &\qquad{\times}\left({R}_x^2{-}\frac{{r}_x^2}{4}\right)e^{-\frac{1}{W_0^2}(|\mathbf{r}_1^\prime|^2+|\mathbf{r}_2^\prime|^2+g^2|\mathbf{r}_3^\prime|^2)}e^{-4i\frac{\Omega}{W_0^2}\mathbf{R}\cdot
  \mathbf{r}^\prime_3}\nonumber\\
 &\qquad{\times}e^{2i\frac{\Omega}{W_0^2}[1{-}\frac{L}{F}]\mathbf{r}^\prime_1\cdot
  \mathbf{r}^\prime_2-2i\frac{\Omega}{W_0^2}\mathbf{r}\cdot\mathbf{r}^\prime_2}\mathcal{J}(\mathbf{r},\mathbf{r}_1^\prime,\mathbf{r}_2^\prime,\mathbf{r}_3^\prime),
\end{align}
with
\begin{align}\label{Jdefinit}
 &\mathcal{J}(\mathbf{r},\mathbf{r}_1^\prime,\mathbf{r}_2^\prime,\mathbf{r}_3^\prime)\\
 &{=}\exp\Bigl[\rho_0^{-\frac{5}{3}}\int\limits_0^1\D\xi\sum\limits_{j=1,2}\Bigl(\left|\mathbf{r}\xi{+}[\mathbf{r}_1^\prime{+}(-1)^j\mathbf{r}_2^\prime](1{-}\xi)\right|^{\frac{5}{3}}\nonumber\\
 &{-\left|\mathbf{r}\xi{+}[\mathbf{r}_1^\prime{+}(-1)^j\mathbf{r}_3^\prime](1{-}\xi)\right|^{\frac{5}{3}}}{-}(1{-}\xi)^{\frac{5}{3}}\left|\mathbf{r}_2^\prime{+}({-}1)^j\mathbf{r}_3^\prime\right|^{\frac{5}{3}}\Bigr)\Bigr],\nonumber
\end{align}
where we have used the variables $\mathbf{r}{=}\mathbf{r}_1{-}\mathbf{r}_2$ and 
$\mathbf{R}{=}(\mathbf{r}_1{+}\mathbf{r}_2)/2$. We integrate over the variables $\mathbf{R}$ and 
$\mathbf{r}_3^\prime$ using the properties of Dirac delta function, which occurs in the integral 
representation of Eq.~(\ref{InitialBW}). For example one can show that
\begin{align}
 \int_{\mathbb{R}^4}\D^2\mathbf{R}\,\D^2\mathbf{r}_3^\prime\,\mathbf{R}^2\,&e^{-4i\frac{\Omega}{W_0^2}\mathbf{R}\cdot\mathbf{r}_3^\prime}f(\mathbf{r}_3^\prime)\nonumber\\
 &=-\frac{(2\pi)^2W_0^8}{(4\Omega)^4}\Delta_{\mathbf{r}_3^\prime}^2 
 f(\mathbf{r}_3^\prime)\Bigl|_{\mathbf{r}_3^\prime=0},
 \label{IntTrick}
\end{align}
where $\Delta_{\mathbf{r}_3^\prime}^2$ is the transverse Laplace operator and  $f(x)$ is an arbitrary 
function. We arrive at
\begin{align}
 &\langle 
 x_0^2\rangle=\frac{2\Omega^2}{(2\pi)^3W_0^6}\int_{\mathbb{R}^{6}}\D^2\mathbf{r}\,\D^2\mathbf{r}_1^\prime\,\D^2\mathbf{r}_2^\prime\left(\frac{g^2
  W_0^2}{2\Omega^2}{-}r_x^2\right)\nonumber\\
 &\qquad{\times}e^{-\frac{1}{W_0^2}(|\mathbf{r}_1^\prime|^2+|\mathbf{r}_2^\prime|^2)}e^{2i\frac{\Omega}{W_0^2}[1{-}\frac{L}{F}]\mathbf{r}^\prime_1\cdot
  \mathbf{r}^\prime_2-2i\frac{\Omega}{W_0^2}\mathbf{r}\cdot\mathbf{r}^\prime_2}\nonumber\\
 &\qquad{\times}\exp\Bigl[\rho_0^{-\frac{5}{3}}\int\limits_0^1\D\xi\Bigl(\sum\limits_{j=1,2}\left|\mathbf{r}\xi{+}[\mathbf{r}_1^\prime{+}(-1)^j\mathbf{r}_2^\prime](1{-}\xi)\right|^{\frac{5}{3}}\nonumber\\
 &\qquad{-2\left|\mathbf{r}\xi{+}\mathbf{r}_1^\prime(1{-}\xi)\right|^{\frac{5}{3}}}{-}2(1{-}\xi)^{\frac{5}{3}}\left|\mathbf{r}_2^\prime\right|^{\frac{5}{3}}\Bigr)\Bigr].\label{BwandGen}
\end{align}
Let us consider the cases of weak and strong turbulence  separately.

\subsection{Weak turbulence}

The weak turbulence is characterized by large values of the parameter $\rho_0$, 
cf.~Eq.~(\ref{rhoDefinition}) together with the dependence on the Rytov parameter in 
(\ref{sigmaRytov}). Hence, we can expand the last exponent of  (\ref{BwandGen}) into series with 
respect to $\rho_0^{-\frac{5}{3}}$ up to the first order. The first term of the expansion which is 
independent of $\rho_0$ in (\ref{BwandGen}), vanishes  and we obtain
\begin{align}
 &\langle 
 x_0^2\rangle=\frac{2\Omega^2\rho_0^{-\frac{5}{3}}}{(2\pi)^3W_0^6}\int_{\mathbb{R}^{6}}\D^2\mathbf{r}\,\D^2\mathbf{r}_1^\prime\,\D^2\mathbf{r}_2^\prime\left(\frac{g^2
  W_0^2}{2\Omega^2}{-}{r}_x^2\right)\nonumber\\
 &\qquad{\times}e^{-\frac{1}{W_0^2}(|\mathbf{r}_1^\prime|^2+|\mathbf{r}_2^\prime|^2)}e^{2i\frac{\Omega}{W_0^2}[1{-}\frac{L}{F}]\mathbf{r}^\prime_1\cdot
  \mathbf{r}^\prime_2-2i\frac{\Omega}{W_0^2}\mathbf{r}\cdot\mathbf{r}^\prime_2}\nonumber\\
 &\qquad{\times}\int\limits_0^1\D\xi\Bigl(\sum\limits_{j=1,2}\left|\mathbf{r}\xi{+}[\mathbf{r}_1^\prime{+}(-1)^j\mathbf{r}_2^\prime](1{-}\xi)\right|^{\frac{5}{3}}\nonumber\\
 &\qquad\quad{-2\left|\mathbf{r}\xi{+}\mathbf{r}_1^\prime(1{-}\xi)\right|^{\frac{5}{3}}}{-}2(1{-}\xi)^{\frac{5}{3}}\left|\mathbf{r}_2^\prime\right|^{\frac{5}{3}}\Bigr).
\end{align}
Performing the multiple integrations in this equation, 
one derives for the beam-wandering variance for a focused beam, $L{=}F$ (defined in 
Eq.~(\ref{Eq:BoundaryConditions})), for weak turbulence the result
\begin{align}
 \langle x_0^2\rangle&=0.94 C_n^2 L^3 W_0^{-\frac{1}{3}}=0.33 W_0^2\sigma_R^2 
 \Omega^{-\frac{7}{6}},\label{bwweak}
\end{align}
where
\begin{align}\label{sigmaRytov}
 \sigma_R^2=1.23 
 C_n^2k^{\frac{7}{6}}L^{\frac{11}{6}}=0.82\rho_0^{-\frac{5}{3}}k^{-\frac{5}{6}}L^{\frac{5}{6}}
\end{align}
is the Rytov parameter \cite{Tatarskii}.

\subsection{Strong turbulence}
For the case of strong turbulence the parameter $\rho_0$ is small. The exponential in 
Eq.~(\ref{Jdefinit}), 
$\mathcal{J}(\mathbf{r},\mathbf{r}_1^\prime,\mathbf{r}_2^\prime,\mathbf{r}_3^\prime)$, significantly 
differs from zero in the following regions:
\begin{align}
 |\mathbf{r}_2^\prime|(1{-}\xi)\gg\rho_0,\quad 
 |\mathbf{r}_3^\prime|(1{-}\xi),\,\,|\mathbf{r}\xi{+}\mathbf{r}_1^\prime(1{-}\xi)|\lesssim\rho_0; 
 \label{region1}
\end{align}
\begin{align}
 |\mathbf{r}\xi{+}\mathbf{r}_1^\prime(1{-}\xi)|\gg\rho_0,\quad|\mathbf{r}_2^\prime|(1{-}\xi),\,|\mathbf{r}_3^\prime|(1{-}\xi)\lesssim\rho_0;
  \label{region2}
 \end{align}
 \begin{align}
 |\mathbf{r}_2^\prime|(1{-}\xi),\,\, 
 |\mathbf{r}_3^\prime|(1{-}\xi),\,\,|\mathbf{r}\xi{+}\mathbf{r}_1^\prime(1{-}\xi)|\lesssim\rho_0.\label{regions}
\end{align}
This function is negligibly small  provided that any of the conditions
\begin{align}
 &|\mathbf{r}_3^\prime|(1{-}\xi)\gg\rho_0,\,\,|\mathbf{r}_2^\prime|(1{-}\xi),\,\,|\mathbf{r}\xi{+}\mathbf{r}_1^\prime(1{-}\xi)|\lesssim\rho_0;\nonumber\\
 &|\mathbf{r}\xi{+}\mathbf{r}_1^\prime(1{-}\xi)|,\,\,|\mathbf{r}_2^\prime|(1{-}\xi)\gg\rho_0,\,\,\,|\mathbf{r}_3^\prime|(1{-}\xi)\lesssim\rho_0;\nonumber\\
 &|\mathbf{r}\xi{+}\mathbf{r}_1^\prime(1{-}\xi)|,\,\,|\mathbf{r}_3^\prime|(1{-}\xi)\gg\rho_0,\,\,|\mathbf{r}_2^\prime|(1{-}\xi)\lesssim\rho_0;\\
 &|\mathbf{r}_2^\prime|(1{-}\xi),\,\,|\mathbf{r}_3^\prime|(1{-}\xi)\gg\rho_0,\,\,|\mathbf{r}\xi{+}\mathbf{r}_1^\prime(1{-}\xi)|\lesssim\rho_0;\nonumber\\
 &|\mathbf{r}\xi{+}\mathbf{r}_1^\prime(1{-}\xi)|,\,\,|\mathbf{r}_2^\prime|(1{-}\xi),\,\,|\mathbf{r}_3^\prime|(1{-}\xi)\gg\rho_0\nonumber
\end{align}
holds true. The function  (\ref{Jdefinit}) can be approximated then as
%
%
%
%
\begin{widetext}
\begin{align}\label{expans}
&\mathcal{J}(\mathbf{r},\mathbf{r}_1^\prime,\mathbf{r}_2^\prime,\mathbf{r}_3^\prime)=\exp\Bigl[-\rho_0^{-\frac{5}{3}}\int\limits_0^1\!\!\D\xi\sum\limits_{j=1,2}\left|\mathbf{r}\xi{+}[\mathbf{r}_1^\prime{+}(-1)^j\mathbf{r}_3^\prime](1{-}\xi)\right|^{\frac{5}{3}}\Bigr]
\sum\limits_{n=0}^\infty\frac{\rho_0^{-\frac{5}{3}n}}{n!}\Bigl\{\sum\limits_{j=1,2}\Bigl(
\int\limits_0^1\!\D\xi\left|\mathbf{r}\xi{+}[\mathbf{r}_1^\prime{+}(-1)^j\mathbf{r}_2^\prime](1-\xi)\right|^{\frac{5}{3}}
\nonumber\\
&-\frac{3}{8}\left|\mathbf{r}_2^\prime{+}(-1)^j\mathbf{r}_3^\prime\right|^{\frac{5}{3}}\Bigr)\Bigr\}^n+\exp\Bigl[-\frac{3}{8}\rho_0^{-\frac{5}{3}}\left|\mathbf{r}_2^\prime{+}(-1)^j\mathbf{r}_3^\prime\right|^{\frac{5}{3}}\Bigr]\sum\limits_{n=0}^\infty\frac{\rho_0^{-\frac{5}{3}n}}{n!}\Bigl\{\sum\limits_{j=1,2}\int\limits_0^1\D\xi\Bigl(\left|\mathbf{r}\xi{+}[\mathbf{r}_1^\prime{+}(-1)^j\mathbf{r}_2^\prime](1-\xi)\right|^{\frac{5}{3}}\\
&-\left|\mathbf{r}\xi{+}[\mathbf{r}_1^\prime{+}(-1)^j\mathbf{r}_3^\prime](1-\xi)\right|^{\frac{5}{3}}\Bigr)\Bigr\}^n-\exp\Bigl[-\rho_0^{-\frac{5}{3}}\sum\limits_{j=1,2}\Bigl\{\frac{3}{8}\left|\mathbf{r}_2^\prime{+}(-1)^j\mathbf{r}_3^\prime\right|^{\frac{5}{3}}+\int\limits_0^1\D\xi\left|\mathbf{r}\xi{+}[\mathbf{r}_1^\prime{+}(-1)^j\mathbf{r}_3^\prime](1-\xi)\right|^{\frac{5}{3}}\Bigr\}\Bigr]\nonumber\\
&\qquad\qquad\qquad\qquad\qquad\qquad\qquad\qquad\times\sum\limits_{n=0}^\infty\frac{\left(\frac{3}{8}\right)^n}{n!}\rho_0^{-\frac{5}{3}n}\Bigl\{\sum\limits_{j=1,2}\int\limits_0^1\D\xi\left|\mathbf{r}\xi{+}[\mathbf{r}_1^\prime{+}(-1)^j\mathbf{r}_2^\prime](1-\xi)\right|^{\frac{5}{3}}\Bigr\}^n\nonumber
\end{align}
 \end{widetext}
Here the first term on the right hand side  accounts for the contributions from the regions 
(\ref{region1}) and (\ref{region2}). If one substitutes the latter into  Eqs.~(\ref{InitialBW}) and 
(\ref{Jdefinit}) and performs integrations, then the  region (\ref{regions}) would be  counted twice. 
Therefore, the last term on the right hand side of  (\ref{expans}) is introduced to eliminate the 
aforementioned double-counting.
It is worth to mention that already the first ($n=0,1$) terms of expansion (\ref{expans}) give a good 
approximation of the function $\mathcal{J}$, cf. Ref.~\cite{Banakh}.

Substituting the right-hand side of Eq.~(\ref{expans})  into Eq.~(\ref{InitialBW}) and integrating over 
the variables $\mathbf{R}$, $\mathbf{r}_3^\prime$ as it is described above, we obtain
\begin{widetext}
\begin{align}\label{longExpr}
& \left\langle 
x_0^2\right\rangle=\frac{2\Omega^2}{(2\pi)^3W_0^6}\int_{\mathbb{R}^{6}}\D^2\mathbf{r}\,\D^2\mathbf{r}_1^\prime\,\D^2\mathbf{r}_2^\prime\left(\frac{g^2
W_0^2}{2\Omega^2}{-}{r}_x^2\right)e^{-\frac{1}{W_0^2}(|\mathbf{r}_1^\prime|^2+|\mathbf{r}_2^\prime|^2)}e^{2i\frac{\Omega}{W_0^2}[1{-}\frac{L}{F}]\mathbf{r}^\prime_1\cdot
  \mathbf{r}^\prime_2-2i\frac{\Omega}{W_0^2}\mathbf{r}\cdot\mathbf{r}^\prime_2}\\
 &{\times}\Biggl\{\exp\Bigl[-\rho_0^{-\frac{5}{3}}\int\limits_0^1\D\xi\sum\limits_{j=1,2}\left|\mathbf{r}\xi{+}[\mathbf{r}_1^\prime{+}(-1)^j\mathbf{r}_3^\prime](1-\xi)\right|^{\frac{5}{3}}\Bigr]\Bigl(1-\frac{3}{4}\rho_0^{-\frac{5}{3}}\left|\mathbf{r}_2^\prime\right|^{\frac{5}{3}}+\rho_0^{-\frac{5}{3}}\sum\limits_{j=1,2}\int\limits_0^1\D\xi\left|\mathbf{r}\xi+[\mathbf{r}_1^\prime{+}({-}1)^j\mathbf{r}_2^\prime](1{-}\xi)\right|^{\frac{5}{3}}\Bigr)\nonumber\\
 &+\exp\Bigl[-\frac{3}{4}\rho_0^{-\frac{5}{3}}|\mathbf{r}_2^\prime|^{\frac{5}{3}}\Bigr]\Bigl(1-2\rho_0^{-\frac{5}{3}}\int\limits_0^1\D\xi\left|\mathbf{r}\xi{+}\mathbf{r}_1^\prime(1-\xi)\right|^{\frac{5}{3}}+\rho_0^{-\frac{5}{3}}\int\limits_0^1\D\xi\sum\limits_{j=1,2}\left|\mathbf{r}\xi{+}[\mathbf{r}_1^\prime{+}(-1)^j\mathbf{r}_2^\prime](1-\xi)\right|^{\frac{5}{3}}\Bigr)\nonumber\\
 &-\exp\Bigl[-\rho_0^{-\frac{5}{3}}\Bigl(\frac{3}{4}\left|\mathbf{r}_2^\prime\right|^{\frac{5}{3}}+2\int\limits_0^1\D\xi\,\left|\mathbf{r}\xi{+}\mathbf{r}_1^\prime(1{-}\xi)\right|^{\frac{5}{3}}\Bigr)\Bigr]\Bigl(1+\rho_0^{-\frac{5}{3}}\int\limits_0^1\D\xi\sum\limits_{j=1,2}\left|\mathbf{r}\xi{+}[\mathbf{r}_1^\prime{+}(-1)^j\mathbf{r}_2^\prime](1-\xi)\right|^{\frac{5}{3}}\Bigr)
 \Biggr\}
 \nonumber
\end{align}
\end{widetext}
The evaluation of  Eq.~(\ref{longExpr}) is simplified further with the use of the approximation 
\cite{Andrews2}
\begin{align}
 \exp\left[-\left(\frac{|\mathbf r|}{\rho_0\Omega}\right)^{\frac{5}{3}}\right]= 
 \exp\left[-\left(\frac{|\mathbf r|}{\rho_0\Omega}\right)^{2}\right],\label{Eq:approx}
\end{align}
which gives good accuracy for small values of $\rho_0$, cf.Ref.~\cite{Mironov2}.

Consecutive integration of (\ref{longExpr}) yields for the  collimated beam 
(\ref{Eq:BoundaryConditions}) with  $F\gg L$ the following result
\begin{align}
 \langle x_0^2\rangle{=}1.78 C_n^{\frac{8}{5}}L^{\frac{37}{15}}k^{-\frac{1}{15}}
 =0.75 W_0^2\sigma_R^{\frac{8}{5}}\Omega^{-1}.\label{sigmarST}
\end{align}
A similar  expression has been obtained by using the Markovian-random-process approximation, cf.  
Ref.~\cite{Mironov2} and the references therein.


\section{Beam-shape distortion}
\label{Sec:BeamShape}

In this Section we derive the expressions for the moments $\langle W_{1/2}^2\rangle$, $\langle 
W_{1/2}^4\rangle$ and $\langle W_1^2W_2^2\rangle$ for weak and strong turbulence regimes. From 
Eqs.~(\ref{Eq:W12EqSxxyy}), (\ref{Eq:SxxViaGamma2})-(\ref{Eq:W1W2ViaGamma}) one can see that 
these moments are expressed through the integrals containing the  field correlation functions 
$\Gamma_2$ and $\Gamma_4$. The first moments of $W_{1/2}^2$ defined by 
Eqs.~(\ref{Eq:SxxMean}),  (\ref{Eq:SyyMean}), and (\ref{Eq:W12EqSxxyy}) contain the following 
integral 
\begin{align}
& \int_{\mathbb R^2}\D^2 \mathbf{r} x^2 \Gamma_2(\mathbf{r})=\frac{ W_0^2}{ \pi^2 
\Omega^4}\int_{\mathbb R^4}\D^2\mathbf{r} \D^2\mathbf{r}^\prime\, 
x^2e^{-\frac{g^2}{2\Omega^2}|\mathbf{r}^\prime|^2}\nonumber\\
 &{\times} \exp\Bigl[{-}\frac{2i}{\Omega}\, \mathbf{r}{\cdot} 
 \mathbf{r}^\prime{-}\rho_0^{-\frac{5}{3}}W_0^{\frac{5}{3}}\int\limits_0^1\D\xi(1{-}\xi)^{\frac{5}{3}}\Bigl(\frac{|\mathbf{r}^\prime|}{\Omega}\Bigr)^{\frac{5}{3}}\Bigr]
 .\label{BeamWidth}
\end{align}
Here we have used Eqs.~(\ref{Gamma2}) and (\ref{DsSph1}).  The second moments of $W_{1/2}^2$ 
defined in Eqs.~(\ref{App:Eq:W12ViaGamma}), (\ref{Eq:W1W2ViaGamma}) contain the   integrals
\begin{align}
 &\int_{\mathbb R^4} \D^2 \mathbf{r}_1\,\D^2 \mathbf{r}_2 x_1^2 x_2^2 
 \Gamma_4(\mathbf{r}_1,\mathbf{r}_2)=\frac{\Omega^2}{2(2\pi)^3  W_0^6}\int_{\mathbb R^6} \D^2 
 \mathbf{r}\, \D^2 \mathbf{r}_1^\prime \D^2 \mathbf{r}_2^\prime\nonumber\\
 &\quad\times  \left(\frac{3g^4W_0^4}{4\Omega^4}-\frac{g^2W_0^2}{\Omega^2}r_x^2+r_x^4\right) 
 e^{-\frac{1}{W_0^2}\left(|\mathbf{r}_1^\prime|^2+|\mathbf{r}_2^\prime|^2\right)}\nonumber\\
 &\quad\times 
 \exp\left[2i\frac{\Omega}{W_0^2}\Bigl(1-\frac{L}{F}\Bigr)\mathbf{r}_1^\prime{\cdot}\mathbf{r}_2^\prime-2i\frac{\Omega}{W_0^2}\mathbf{r}{\cdot}\mathbf{r}_2^\prime\right]
  \label{beamWIntG4}\\
&\qquad{\times}\exp\Bigl[\rho_0^{-\frac{5}{3}}\int\limits_0^1\D\xi\Bigl(\sum\limits_{j=1,2}
\left|\mathbf{r}\xi{+}[\mathbf{r}_1^\prime{+}(-1)^j\mathbf{r}_2^\prime](1{-}\xi)\right|^{\frac{5}{3}}\nonumber\\
 &\qquad\qquad{-2\left|\mathbf{r}\xi{+}\mathbf{r}_1^\prime(1{-}\xi)\right|^{\frac{5}{3}}}{-}2(1{-}\xi)^{\frac{5}{3}}
 \left|\mathbf{r}_2^\prime\right|^{\frac{5}{3}}\Bigr)\Bigr]\nonumber
\end{align}
and
\begin{align}
 &\int_{\mathbb R^4} \D^2 \mathbf{r}_1\,\D^2 \mathbf{r}_2 x_1^2 y_2^2 
 \Gamma_4(\mathbf{r}_1,\mathbf{r}_2)=\frac{\Omega^2}{2(2\pi)^3  W_0^6}\int_{\mathbb R^6} \D^2 
 \mathbf{r}\, \D^2 \mathbf{r}_1^\prime \D^2 \mathbf{r}_2^\prime\nonumber\\
 &\quad\times  
 \left(\frac{g^4W_0^4}{4\Omega^4}+\frac{g^2W_0^2}{\Omega^2}r_x^2+r_x^2r_y^2\right) 
 e^{-\frac{1}{W_0^2}\left(|\mathbf{r}_1^\prime|^2+|\mathbf{r}_2^\prime|^2\right)}\nonumber\\
 &\quad\times 
 \exp\left[2i\frac{\Omega}{W_0^2}\Bigl(1-\frac{L}{F}\Bigr)\mathbf{r}_1^\prime{\cdot}\mathbf{r}_2^\prime-2i\frac{\Omega}{W_0^2}\mathbf{r}{\cdot}\mathbf{r}_2^\prime\right]
  \label{beamCorrIntG4}\\
&\qquad{\times}\exp\Bigl[\rho_0^{-\frac{5}{3}}\int\limits_0^1\D\xi\Bigl(\sum\limits_{j=1,2}
\left|\mathbf{r}\xi{+}[\mathbf{r}_1^\prime{+}(-1)^j\mathbf{r}_2^\prime](1{-}\xi)\right|^{\frac{5}{3}}\nonumber\\
 &\qquad\qquad{-2\left|\mathbf{r}\xi{+}\mathbf{r}_1^\prime(1{-}\xi)\right|^{\frac{5}{3}}}{-}2(1{-}\xi)^{\frac{5}{3}}
 \left|\mathbf{r}_2^\prime\right|^{\frac{5}{3}}\Bigr)\Bigr].\nonumber
\end{align}
Here we have used the definition of $\Gamma_4$ given in Eq.~(\ref{Gamma4}) and performed the 
four-fold integration 
in a similar  way as  in Eq.~(\ref{InitialBW}), with the aid of formulas similar to (\ref{IntTrick}).

\subsection{Weak turbulence}
In the limit of weak turbulence we derive, by substituting Eqs.~(\ref{BeamWidth}), (\ref{bwweak}) in 
Eqs.~(\ref{Eq:SxxMean}) and (\ref{Eq:W12EqSxxyy}), the following result for the first moment of 
$W_{1/2}^2$: 
 \begin{align}
  &\langle W_{1/2}^2\rangle{=}
  \frac{W_0^2}{\Omega^2}{+}2.96 W_0^2 \sigma_R^2\Omega^{-\frac{7}{6}}.\label{IntWT}
 \end{align}
 In Eq.~(\ref{BeamWidth}) we have used the approximations 
$\bigl(|\mathbf{r}^\prime|/\Omega\bigr)^\frac{5}{3}\approx 
\bigl(|\mathbf{r}^\prime|/\Omega\bigr)^2$, cf.~\cite{Andrews2} and 
$\int_0^1\D\xi f(\xi)\approx f(0)$, cf.~\cite{Kon}.  The first term in Eq.~(\ref{IntWT})  describes the  
diffraction broadening in free space and the second term gives the amount of diffraction broadening 
in turbulence.

The second order moments of $W_{1/2}^2$ are evaluated by substituting Eqs.~(\ref{bwweak}), 
(\ref{beamWIntG4}),  (\ref{IntWT}) in Eq.~(\ref{App:Eq:W12ViaGamma}) and correspondingly 
Eqs.~(\ref{bwweak}), (\ref{beamCorrIntG4}),  (\ref{IntWT}) in Eq.~(\ref{Eq:W1W2ViaGamma}). We 
evaluate the integrals in Eqs.~(\ref{beamWIntG4}) and (\ref{beamCorrIntG4}) by expanding the last 
exponents into series with respect to $\rho_0^{-\frac{5}{3}}$ up to the second order and consecutive 
integration. For a focused beam ($L{=}F$) we obtain
\begin{align}
 &\int_{\mathbb R^4} \D^2 \mathbf{r}_1\,\D^2 \mathbf{r}_2 x_1^2 x_2^2 
 \Gamma_4(\mathbf{r}_1,\mathbf{r}_2)\nonumber\\
 &\quad=\frac{W_0^4}{16 \Omega^4}+0.58 W_0^4\sigma_R^2\Omega^{-\frac{19}{6}}+1.37 
 W_0^4\sigma_R^4\Omega^{-\frac{7}{3}},
\end{align}
\begin{align}
 &\int_{\mathbb R^4} \D^2 \mathbf{r}_1\,\D^2 \mathbf{r}_2 x_1^2 y_2^2 
 \Gamma_4(\mathbf{r}_1,\mathbf{r}_2)\nonumber\\
 &\quad=\frac{W_0^4}{16 \Omega^4}+0.51 W_0^4\sigma_R^2\Omega^{-\frac{19}{6}}+1.145 
 W_0^4\sigma_R^4\Omega^{-\frac{7}{3}}.
\end{align}
The corresponding  (co)variances are evaluated as
\begin{align}
 \left\langle(\Delta W_{1/2}^2)^2\right\rangle=1.2 W_0^4\sigma_R^2\Omega^{-\frac{19}{6}}{+}0.17 
 W_0^4\sigma_R^4\Omega^{-\frac{7}{3}},
\end{align}
\begin{align}
 \left\langle\Delta W_{1}^2\Delta W_2^2\right\rangle{=}{-}0.8 
 W_0^4\sigma_R^2\Omega^{-\frac{19}{6}}{-}0.05 W_0^4\sigma_R^4\Omega^{-\frac{7}{3}},
\end{align}
correspondingly. The correlation function for weak turbulence  is negative, i.e. the shape of the 
ellipse is deformed in such a way that the increase  of the beam width along one  half-axis of the 
ellipse causes the decrease of the width in the complementary direction.

 \subsection{Strong turbulence}
 For strong turbulence the first moment of $W_{1/2}^2$ is evaluated by substituting 
 Eqs.~(\ref{BeamWidth}) and (\ref{sigmarST})
 in Eqs.~(\ref{Eq:SxxMean}), (\ref{Eq:W12EqSxxyy}). We also use the approximation (\ref{Eq:approx}) 
 for evaluating (\ref{BeamWidth}) to obtain
 \begin{align}\label{BeamSpreadST}
 \langle W_{1/2}^2\rangle{=}\gamma W_0^2&+1.71 
 W_0^2\sigma_R^{\frac{12}{5}}\Omega^{{-}1}{-}2.99 W_0^2\sigma_R^{\frac{8}{5}}\Omega^{{-}1},
 \end{align}
 where $\gamma{=}(1{+}\Omega^2)/\Omega^2$. It is also assumed that $\Omega>1$, cf.~ 
 Eq.~(\ref{FresnelOmega}).
 
For calculating the (co)variances of $W_{1/2}^2$ we firstly evaluate the integrals in 
(\ref{beamWIntG4}) and (\ref{beamCorrIntG4}) by using the approximation (\ref{expans}) in the way  
described in Section~\ref{Sec:BW}. Within this approximation one gets, for example
\begin{widetext}
\begin{align}\label{longBBroad}
&\int_{\mathbb R^4} \D^2 \mathbf{r}_1\,\D^2 \mathbf{r}_2 x_1^2 x_2^2 
\Gamma_4(\mathbf{r}_1,\mathbf{r}_2)=\frac{\Omega^2}{2(2\pi)^3W_0^6}\int_{\mathbb{R}^{6}}\D^2\mathbf{r}\,\D^2\mathbf{r}_1^\prime\,\D^2\mathbf{r}_2^\prime\left(\frac{3}{4}\gamma^2W_0^4{-}\gamma
 W_0^2{r}_x^2+r_x^4\right)\nonumber\\
&\qquad\quad\times 
\exp\Bigl[-\frac{1}{W_0^2}(|\mathbf{r}_1^\prime|^2+|\mathbf{r}_2^\prime|^2)+2i\frac{\Omega}{W_0^2}\mathbf{r}^\prime_1\cdot
  \mathbf{r}^\prime_2-2i\frac{\Omega}{W_0^2}\mathbf{r}\cdot\mathbf{r}^\prime_2\Bigr]\\
 &{\times}\Biggl\{\exp\Bigl[-\rho_0^{-\frac{5}{3}}\int\limits_0^1\D\xi\sum\limits_{j=1,2}\left|\mathbf{r}\xi{+}[\mathbf{r}_1^\prime{+}(-1)^j\mathbf{r}_3^\prime](1{-}\xi)\right|^{\frac{5}{3}}\Bigr]\Bigl(1{-}\frac{3}{4}\rho_0^{-\frac{5}{3}}\left|\mathbf{r}_2^\prime\right|^{\frac{5}{3}}{+}\rho_0^{-\frac{5}{3}}\sum\limits_{j=1,2}\int\limits_0^1\D\xi\left|\mathbf{r}\xi+[\mathbf{r}_1^\prime{+}({-}1)^j\mathbf{r}_2^\prime](1{-}\xi)\right|^{\frac{5}{3}}\Bigr)\nonumber\\
 &\qquad+\exp\Bigl[-\frac{3}{4}\rho_0^{-\frac{5}{3}}|\mathbf{r}_2^\prime|^{\frac{5}{3}}\Bigr]\Bigl(1-2\rho_0^{-\frac{5}{3}}\int\limits_0^1\D\xi\left|\mathbf{r}\xi{+}\mathbf{r}_1^\prime(1{-}\xi)\right|^{\frac{5}{3}}+\rho_0^{-\frac{5}{3}}\int\limits_0^1\D\xi\sum\limits_{j=1,2}\left|\mathbf{r}\xi{+}[\mathbf{r}_1^\prime{+}(-1)^j\mathbf{r}_2^\prime](1{-}\xi)\right|^{\frac{5}{3}}\Bigr)\nonumber\\
 &\qquad-\exp\Bigl[-\rho_0^{-\frac{5}{3}}\Bigl(\frac{3}{4}\left|\mathbf{r}_2^\prime\right|^{\frac{5}{3}}{+}2\int\limits_0^1\D\xi\,\left|\mathbf{r}\xi{+}\mathbf{r}_1^\prime(1{-}\xi)\right|^{\frac{5}{3}}\Bigr)\Bigr]\Bigl(1+\rho_0^{-\frac{5}{3}}\int\limits_0^1\D\xi\sum\limits_{j=1,2}\left|\mathbf{r}\xi{+}[\mathbf{r}_1^\prime{+}(-1)^j\mathbf{r}_2^\prime](1{-}\xi)\right|^{\frac{5}{3}}\Bigr)
 \Biggr\}.
 \nonumber
\end{align}
\end{widetext}
Performing the multiple integration in (\ref{longBBroad}), one derives
   \begin{align}
   \int_{\mathbb R^4} \D^2 \mathbf{r}_1\,\D^2 \mathbf{r}_2 x_1^2& x_2^2 
   \Gamma_4(\mathbf{r}_1,\mathbf{r}_2)\nonumber\\
   &=\gamma^2\frac{W_0^4}{16}+4.34 \gamma W_0^4 \sigma_R^{\frac{12}{5}}\Omega^{-1}
   \label{GammaStr2}
 \end{align}
 and similarly
  \begin{align}
   \int_{\mathbb R^4} \D^2 \mathbf{r}_1\,\D^2 \mathbf{r}_2 &x_1^2 y_2^2 
   \Gamma_4(\mathbf{r}_1,\mathbf{r}_2)\nonumber\\
   &=\gamma^2\frac{W_0^4}{16}+3.16 \gamma W_0^4 \sigma_R^{\frac{12}{5}}\Omega^{-1}.
   \label{GammaStr3}
  \end{align}
 Finally, substituting Eqs.~(\ref{sigmarST}), (\ref{BeamSpreadST}), (\ref{GammaStr2}) and 
 (\ref{GammaStr3}) into Eqs.~(\ref{App:Eq:W12ViaGamma}) and (\ref{Eq:W1W2ViaGamma}) we 
 obtain
 \begin{align}
  \left\langle (\Delta W_{1/2}^2)^2\right\rangle=13.14 \gamma 
  W_0^4\sigma_R^{\frac{12}{5}}\Omega^{-1}
 \end{align}
and
 \begin{align}
  \left\langle \Delta W_{1}^2\Delta W_{2}^2\right\rangle=0.65\gamma 
  W_0^4\sigma_R^{\frac{12}{5}}\Omega^{-1}.
  \label{CovarianceST}
 \end{align}
It is worth to note that in contrast to weak turbulence case the covariance (\ref{CovarianceST}) is 
positive, i.e. the shape of the beam profile of the ellipse is deformed in such a way that the increase 
of beam width along one  half-axis of the ellipse causes the increase in the complimentary direction.   
 
%

\section{Mean values and covariance matrix elements}\label{Sec:CovMatrixEl}
\begin{table}[h!]
\caption{Mean values and  elements of the covariance matrix of the vector $\mathbf{v}$, are given
for horizontal links, in terms of the
transmitter beam spot radius, $W_0$, the Fresnel parameter of the beam,
$\Omega{=}\frac{kW_0^2}{2L}$, and the Rytov parameter,
 $\sigma_R^2=1.23 C_n^2\,k^{\frac{7}{6}}L^{\frac{11}{6}}$. Here $k$ is the beam wave-number, $L$ 
 is
the propagation distance,
 $C_n^2$ $[m^{-\frac{2}{3}}]$ is the   structure constant of the
 refractive index of the air, and
 $\gamma{=}(1{+}\Omega^2)/\Omega^2$.}
\renewcommand{\arraystretch}{2.5}
\begin{tabular}{cc}
\hline
\hline
&
\multicolumn{1}{c}{Weak turbulence} \\
\hline
$\left\langle \Theta_{1/2}\right\rangle$&\quad  $\ln\Biggl[\frac{\left(1+2.96 
\sigma_R^2\Omega^{\frac{5}{6}}\right)^2}{\Omega^2\sqrt{\left(1+2.96 
\sigma_R^2\Omega^{\frac{5}{6}}\right)^2+1.2\sigma_R^2\Omega^{\frac{5}{6}}}}\Biggr]$\\
$\left\langle\Delta x_0^2\right\rangle,\left\langle\Delta y_0^2\right\rangle$&$0.33\,W_0^2 
\sigma_R^2 \Omega^{-\frac{7}{6}}$ \\
$\left\langle \Delta \Theta_{1/2}^2\right\rangle$
&$\ln\Biggl[1+\frac{1.2\sigma_R^2\Omega^{\frac{5}{6}}}{\left(1+2.96\sigma_R^2\Omega^{\frac{5}{6}}\right)^2}\Biggr]$
 \\
$\left\langle \Delta \Theta_1\Delta 
\Theta_2\right\rangle$&$\ln\Biggl[1-\frac{0.8\sigma_R^2\Omega^{\frac{5}{6}}}{\left(1+2.96\sigma_R^2\Omega^{\frac{5}{6}}\right)^2}\Biggr]$
 \\
\hline
\hline
&
\multicolumn{1}{c}{Strong turbulence} \\
\hline
$\left\langle \Theta_{1/2}\right\rangle$&  $\ln\Biggl[\frac{\bigl(\gamma+1.71 
\sigma_R^{\frac{12}{5}}\Omega^{{-}1}{-}2.99 
\sigma_R^{\frac{8}{5}}\Omega^{{-}1}\bigr)^2}{\sqrt{\bigl(\gamma+1.71 
\sigma_R^{\frac{12}{5}}\Omega^{{-}1}{-}2.99 \sigma_R^{\frac{8}{5}}\Omega^{{-}1}\bigr)^2+3.24 
\gamma\sigma_R^{\frac{12}{5}}\Omega^{-1}}}\Biggr]$\\
$\left\langle\Delta x_0^2\right\rangle,\left\langle\Delta y_0^2\right\rangle$&$0.75\,W_0^2 
\sigma_R^{\frac{8}{5}} \Omega^{-1}$ \\
$\left\langle \Delta \Theta_{1/2}^2\right\rangle$
&$\ln\Biggl[1+\frac{13.14 \gamma\sigma_R^{12/5}\Omega^{-1}}{\bigl(\gamma+1.71 
\sigma_R^{\frac{12}{5}}\Omega^{{-}1}{-}2.99 \sigma_R^{\frac{8}{5}}\Omega^{{-}1}\bigr)^2}\Biggr]$ \\
$\left\langle \Delta \Theta_1\Delta \Theta_2\right\rangle$&$\ln\Biggl[1+\frac{0.65\gamma 
\sigma_R^{12/5}\Omega^{-1}}{\bigl(\gamma+1.71 \sigma_R^{\frac{12}{5}}\Omega^{{-}1}{-}2.99 
\sigma_R^{\frac{8}{5}}\Omega^{{-}1}\bigr)^2}\Biggr]$ \\
\hline
\hline
\end{tabular}
\label{tab:covariance}
\end{table}
The Table~\ref{tab:covariance} lists the non-zero means and covariance matrix elements of the 
four-dimensional Gaussian distribution for the random vector $\mathbf{v}$  defined in 
Eq.~(\ref{Eq:vMultNoise}).   We list the results for weak and strong turbulence.
The weak turbulence results can be applied, e.g., for short propagation distances with $\sigma_R^2 
\lesssim 1$. In near-to-ground propagation the latter condition is fulfilled for optical frequencies for  
night-time communication. The strong turbulence results are applied for  short distance 
communication,   $\sigma_R^2\gg 1$. For a near-to-ground communication scenario this 
corresponds to the day-time operation at clear sunny days.

\section{Log-normal model}\label{Sec:LogNormal}

 The log-normal  probability distribution for transmittance is 
\begin{align}
 \mathcal{P}(\eta)=\frac{1}{\eta\sigma\sqrt{2\pi}}\exp\left[-\frac{\Bigl(-\ln\eta-\mu\Bigr)^2}{2\sigma^2}\right]
\end{align}
where 
\begin{align}
 \mu=-\ln\left(\frac{\langle\eta\rangle^2}{\sqrt{\langle\eta^2\rangle}}\right)
\end{align}
and 
\begin{align}
 \sigma^2=\ln\left(\frac{\langle\eta^2\rangle}{\langle\eta\rangle^2}\right),\label{SigmaLogNormal1}
\end{align}
are parameters of the log-normal distribution. The parameters $\mu$ and $\sigma$ are the functions 
of the first and second moments of transmittance 
\begin{align}
 \left\langle\eta\right\rangle=\int_\mathcal{A}\D^2 \mathbf{r} \Gamma_2(\mathbf{r}),\label{meanEta}
\end{align}
\begin{align}
 \left\langle\eta^2\right\rangle{=}\int_\mathcal{A}\D^2 
 \mathbf{r}_1\D^2\mathbf{r}_2\Gamma_4(\mathbf{r}_1,\mathbf{r}_2),\label{squareEta}
\end{align}
where the field coherence functions $\Gamma_2$ and $\Gamma_4$ are given by 
Eqs.~(\ref{Gamma2}) and (\ref{Gamma4}), respectively. Here the integration is performed over  the 
circular aperture opening area $\mathcal{A}$.

The first moment of transmittance (\ref{meanEta}) is evaluated explicitly as
\begin{align}
\left\langle\eta\right\rangle=1-\exp\left[-\frac{2a^2}{\langle W^2\rangle}\right],\label{MeanEta}
\end{align}
where $a$ is the aperture radius and 
\begin{align}
\langle W^2\rangle=\langle S_{xx}\rangle+4\langle x_0^2\rangle 
\end{align}
is the so called "long-term" beam mean-square radius \cite{Fante1}.  Here $\langle S_{xx}\rangle$ 
and $\langle x_0^2\rangle $ are defined by Eqs.~(\ref{Eq:SxxMean}) and (\ref{App:bwvariance}) 
respectively.
For weak turbulence from Eqs.~(\ref{IntWT}) and (\ref{bwweak}) we evaluate $\langle 
W^2\rangle=W_0^2\Omega^{-2}+4.33W_0^2\sigma_R^2\Omega^{-\frac{7}{6}}$. However, the 
integration of Eq.~(\ref{squareEta}) is more involved. 
In this Letter we evaluated Eq.~(\ref{squareEta}) numerically.

\end{document}